\documentclass[aps, prd, twocolumn, preprintnumbers, groupedaddress, nofootinbib, amssymb, notitlepage, eqsecnum]{revtex4-2}
\usepackage{bm}
\usepackage{amsmath,amsthm,amssymb}
\usepackage{amsfonts}
\usepackage{hyperref}

\allowdisplaybreaks[1]

\newcommand{\rd}{{\rm d}}
\newcommand{\be}{\begin{equation}}
\newcommand{\ee}{\end{equation}}
\newcommand{\ba}{\begin{eqnarray}}
\newcommand{\ea}{\end{eqnarray}}

\begin{document}

\preprint{YITP-24-74, IPMU24-0028, WUCG-24-06}

\title{Even- and odd-parity stabilities of black holes in Einstein-Aether gravity}

\author{Antonio De Felice$^{a}$}
\email{antonio.defelice@yukawa.kyoto-u.ac.jp}  

\author{Shinji Mukohyama$^{a, b}$}
\email{shinji.mukohyama@yukawa.kyoto-u.ac.jp} 

\author{Shinji Tsujikawa$^{c}$}
\email{tsujikawa@waseda.jp} 
 
\author{Anzhong Wang$^{d}$}
\email{anzhong$\_$wang@baylor.edu}

\author{Chao Zhang$^{e,f}$}
\email{phyzc@cup.edu.cn}

\affiliation{$^{a}$Center for Gravitational Physics and Quantum Information, Yukawa Institute for Theoretical Physics, Kyoto University, 606-8502, Kyoto, Japan\\
$^{b}$Kavli Institute for the Physics and Mathematics of the Universe (WPI),
The University of Tokyo Institutes for Advanced Study,
The University of Tokyo, Kashiwa, Chiba 277-8583, Japan\\
$^{c}$Department of Physics, Waseda University, 3-4-1 Okubo, 
Shinjuku, Tokyo 169-8555, Japan\\
$^{d}$GCAP-CASPER, Physics Department, Baylor
University, Waco, TX 76798-7316, USA\\
$^{e}$Basic Research Center for Energy Interdisciplinary, College of Science, China University of Petroleum-Beijing, Beijing 102249, China\\
$^{f}$Beijing Key Laboratory of Optical Detection Technology for Oil and Gas, China University of Petroleum-Beijing, Beijing, 102249, China}
 
\date{\today}

\begin{abstract}

In Einstein-Aether theories with a timelike unit vector field, we study the linear stability of static and spherically symmetric black holes against both even- and odd-parity perturbations. 
For this purpose, we formulate a gauge-invariant black hole perturbation theory in the background Aether-orthogonal frame where the spacelike property of hypersurfaces orthogonal to the timelike Aether field is always maintained even inside the metric horizon. Using a short-wavelength approximation with large radial and angular momenta, we show that, in general, there are three dynamical degrees 
of freedom arising from the even-parity sector besides two propagating degrees of freedom present in the odd-parity sector. The propagation speeds of even-parity perturbations and their no-ghost conditions coincide with those of tensor, vector, and scalar perturbations on the Minkowski background, while the odd sector contains tensor and vector modes with the same propagation speeds as those in the even-parity sector (and hence as those on the Minkowski background).
Thus, the consistent study of black hole perturbations in the Aether-orthogonal frame on static and spherically symmetric backgrounds does not add new small-scale stability conditions to those known  for the Minkowski background in the literature.

\end{abstract}


\maketitle

\section{Introduction}
\label{Intro}

Gravity has been the most difficult force in Nature to bring under theoretical control, both at the classical and quantum levels. 
In General Relativity (GR), the presence of a dimensionful gravitational coupling constant, $G=m_{\rm pl}^{-2}$, leads to the 
non-renormalizability as a quantum field theory in four spacetime dimensions. 
One possible way to evade this problem is to break the Lorentz symmetry at high energy \cite{Chadha:1982qq, Kostelecky:1988zi, Gambini:1998it, Douglas:2001ba, Carroll:2001ws, Amelino-Camelia:2008aez}. 
A well-known example of this type is Ho$\check{\rm r}$ava-Lifshitz gravity \cite{Horava:2009uw}, in which a power-counting renormalizability can be realized by anisotropic scaling in the $3 + 1$ ADM formalism.

There are  other mechanisms to break the Lorentz invariance and one of them is to invoke a timelike unit vector field $u^{\mu}$ \cite{Gasperini:1987nq, Jacobson:2000xp}. 
The non-vanishing property of such an ``Aether'' field, which can be encoded by a Lagrangian in the form $\lambda (u_{\mu} u^{\mu}+1)$ with $\lambda$ being a Lagrange multiplier, leads to the breaking of Lorentz symmetry at any spacetime point.
The preferred timelike threading with respect to $u^{\mu}$ should be dynamical to maintain the general covariance of the theory. 
Einstein-Aether (EA) theory advocated by Jacobson and Mattingly \cite{Jacobson:2000xp} is characterized by four coupling constants $c_{1,2,3,4}$ of derivative interactions of the vector field besides the Einstein-Hilbert term.

It is known that the EA framework can encompass several different khronometric theories, each of which is characterized by the presence of a locally timelike ``khronon'' field  $\tau$ satisfying $u_{\mu} \propto \partial_{\mu} \tau$, 
as specific cases. 
The khronometric theory corresponding to the infrared limit of the non-projectable version of Ho$\check{\rm r}$ava-Lifshitz gravity \cite{Blas:2009qj} admits all 
vorticity-free solutions of 
EA theory\footnote{\label{footnote:nonprojectableHL}The former admits extra vorticity-free solutions that are not solutions of the latter. They can be eliminated by some boundary conditions. Even after that, the mapping between the two theories is not one-to-one since the vorticity-free solutions of the latter do not depend on the coefficient of the vorticity squared in the action.} \cite{Jacobson:2010mx}.
On the other hand, the khronometric theory corresponding to the infrared limit of the projectable Ho$\check{\rm r}$ava-Lifshitz gravity \cite{Horava:2009uw} admits all vorticity- and acceleration-free solutions of EA theory\footnote{Comments similar to those in footnote \ref{footnote:nonprojectableHL} apply also here.} \cite{Haghani:2014ita,Jacobson:2014mda}.  
In this limit, the dynamics is rather non-trivial as one needs to take into account an analogue of the Vainshtein screening~\cite{Mukohyama:2010xz}.
Provided that vorticity is absent in $u_{\mu}$, EA theory also accommodates cuscuton theory \cite{Afshordi:2006ad} with a quadratic scalar potential for the coefficients 
$c_2\neq 0$ and $c_{1,3,4}=0$ \cite{Afshordi:2006ad, Bhattacharyya:2016mah}. 
We refer the readers to \cite{Gripaios:2004ms, Zlosnik:2006zu, Kanno:2006ty, Zlosnik:2006zu, Chesler:2017khz} for extended versions of EA theory.

For general non-vanishing constants $c_{1,2,3,4}$, EA theory contains two tensor, two vector, and one scalar degrees of freedom (DOFs). 
On the Minkowski background, their propagation speeds ($c_T$, $c_V$, $c_S$, respectively) are constants  expressed in terms of $c_{1,2,3,4}$ alone \cite{Jacobson:2004ts}.
The superluminal propagation of such modes can be allowed without violating causality in theories with broken Lorentz invariance. 
This result is consistent with the experiments of gravitational \v{C}erenkov radiation \cite{Elliott:2005va}, in which the deviation of $c_T$ from 1 in the sub-luminal range ($c_T<1$) is highly restricted. 
The speed of tensor perturbations constrained from gravitational waves emitted from a neutron star binary \cite{LIGOScientific:2017zic} placed the tight limit $|c_1+c_3| \lesssim 10^{-15}$ \cite{Gong:2018cgj, Oost:2018tcv}. 
The coupling constants $c_i$ are further constrained by solar-system tests of gravity \cite{Foster:2005dk}, big-bang nucleosynthesis \cite{Carroll:2004ai}, binary pulsars \cite{Foster:2007gr, Yagi:2013qpa, Yagi:2013ava, Gupta:2021vdj}, and gravitational waveforms \cite{Hansen:2014ewa, Zhang:2019iim, Schumacher:2023cxh}, but there are still allowed parameter spaces consistent with those data.

In EA theory, there are hairy black hole (BH) solutions with a non-vanishing Aether-field profile on  static and spherically symmetric (SSS) geometries \cite{Eling:2006ec, Foster:2005fr, Konoplya:2006rv, Garfinkle:2007bk, Berglund:2012bu, Gao:2013im, Lin:2014eaa, Ding:2015kba, Ding:2016wcf, Ding:2017gfw, Lin:2017cmn, Ding:2018whp, Chan:2019mdn, Zhu:2019ura, Chan:2020amr, Khodadi:2020gns, Zhang:2020too, Oost:2021tqi, Mazza:2023iwv}. 
A striking property of such BHs is the existence of a universal horizon that sets a causal boundary for any speed of propagating DOFs \cite{Blas:2011ni}. 
Since the universal horizon is located inside the metric horizon, there are regions in which the spacelike property of constant $t$ hypersurfaces (where $t$ is a Killing time) does not hold. Upon using the standard SSS coordinate system with the Killing time $t$ and the areal radius $r$, the stability of BHs against linear perturbations is not firmly determined in theories with preferred timelike threading.

The issue of proper linear stability analysis for BHs in EA theory was recently addressed in Ref.~\cite{Mukohyama:2024vsn} for perturbations in the odd-parity sector of  SSS backgrounds.  
For this purpose, it is  important to resort to an Aether-orthogonal frame characterized by a timelike coordinate $\tau$ and a spacelike coordinate $\rho$, in which the timelike Aether field is orthogonal to the spacelike hypersurfaces defined by constant values of $\tau$. Since we are considering  SSS backgrounds without vorticity, the background Aether field $u_{\mu}$ can be related to the Khronon field $\tau$ as $u_{\mu}=-f_A (r) \partial_{\mu} \tau$, where $f_A$ is a function of the areal distance $r$.
Since the second-order action of odd-parity perturbations in the $(t,r)$ coordinates was derived in Ref.~\cite{Tsujikawa:2021typ}, the linear stability conditions in the Aether-orthogonal frame were addressed in Ref.~\cite{Mukohyama:2024vsn} by using transformation properties of two dynamical perturbations. 
It was shown that, in the short-wavelength limit, the radial and angular propagation speeds as well as the no-ghost conditions of odd-parity perturbations are equivalent to those of tensor and vector perturbations on the Minkowski background.

In this paper, we address the linear stability of SSS BHs in EA gravity by considering even-parity perturbations as well as odd-parity modes in the Aether-orthogonal frame.
Due to the complicated structure of the second-order action of even-parity perturbations, this issue was not thoroughly studied, even for the standard $(t,r)$ coordinates. 
Choosing appropriate gauge conditions, the number of dynamical DOFs in the even-parity sector reduces to three for general 
non-vanishing coefficients $c_{1,2,3,4}$.
Taking the small-wavelength approximation for both radial and angular momentum modes, the propagation speeds and no-ghost conditions of the three dynamical perturbations will be shown to coincide with those of tensor, vector, and scalar perturbations on the Minkowski background. We will also study the behavior of odd-parity perturbations directly in the Aether-orthogonal frame and show that the linear stability conditions are equivalent to those derived in Ref.~\cite{Mukohyama:2024vsn} under the coordinate transformation of 
dynamical perturbations, and thus to those for tensor and vector perturbations on the Minkowski background. 
 
This paper is organized as follows. 
In Sec.~\ref{EAsec}, we introduce the Aether-orthogonal frame, where the line element is given by Eq.~(\ref{phipsi}) with 
$f_A^2 \geq 0$ and $h_A^2 \geq 0$.
In Sec.~\ref{stasec}, we address the gauge-invariant formulation of BH perturbations in the Aether-orthogonal frame and discuss several different gauge choices for fixing residual gauge DOFs.
In Sec.~\ref{oddsec}, we expand the action up to  second order in odd-parity perturbations and reproduce the stability results of Ref.~\cite{Mukohyama:2024vsn} derived 
by using a different method.
In Sec.~\ref{evensec}, we explore how the number of dynamical DOFs in the even-parity sector reduces to three under the gauge choice (\ref{gachoice}) and derive the linear stability conditions for modes with large radial and angular momenta. Sec.~\ref{concludesec} is devoted to conclusions.

Throughout this paper, we use the natural units in which the speed of light $c$ and the reduced Planck constant $\hbar$ are unity. 
We also adopt the metric signature $(-,+,+,+)$. 
The Greek indices run from 0 to 3, while the Latin indices run from 1 to 3.

\section{Aether orthogonal frame}
\label{EAsec}

We consider EA theory given by the action \cite{Jacobson:2000xp}
\be
{\cal S}=\frac{1}{16\pi G_{\ae}} \int  {\rm d}^4 x \sqrt{-g} 
\left[ R+{\cal L}_{\ae}+\lambda (g_{\mu \nu} u^{\mu} u^{\nu}+1) 
\right],
\label{action}
\ee
where $G_{\ae}$ is the gravitational coupling constant of the theory, $g$ is a determinant of the metric tensor $g_{\mu \nu}$, $R$ denotes the Ricci scalar, and $\lambda$ is a Lagrange multiplier.
The Aether field $u_{\mu}$ has derivative couplings of the form 
\be
{\cal L}_{\ae}=-{M^{\alpha \beta}}_{\mu \nu}
\nabla_{\alpha} u^{\mu} \nabla_{\beta} u^{\nu}\,,
\ee
with 
\be
{M^{\alpha \beta}}_{\mu \nu}:=c_1 g^{\alpha \beta} g_{\mu \nu}
+c_2 \delta^{\alpha}_{\mu} \delta^{\beta}_{\nu}
+c_3 \delta^{\alpha}_{\nu} \delta^{\beta}_{\mu}
-c_4 u^{\alpha} u^{\beta} g_{\mu \nu}\,,
\ee
where $\nabla_{\alpha}$ denotes the covariant derivative operator, and $c_{1,2,3,4}$ are the four dimensionless coupling constants. 
Varying the action (\ref{action}) with respect to $\lambda$ leads to  
\be
g_{\mu \nu} u^{\mu} u^{\nu}=-1\,,
\label{be1}
\ee
which ensures the existence of a timelike unit vector field at any spacetime point. 

The SSS background can be described by the line element 
\be
\rd s^2 = -f(r) \rd t^2 + h^{-1}(r)\rd r^2 
+ r^2 \left( \rd \theta^2+\sin^2 \theta\,
\rd\varphi^2 \right)\,, \label{SSS}
\ee
where $t$ is the Killing time, $r$ is the areal distance, and $f, h$ are functions of $r$.
Alternatively, one can also choose the Eddington-Finkelstein coordinate \cite{Zhang:2020too}
\be
\rd s^2 = -f(r) \rd v^2 +2B(r) \rd v \rd r
+ r^2 \left( \rd \theta^2+\sin^2 \theta\,
\rd\varphi^2 \right), \label{SSS2}
\ee
where $v$ and $B(r)$ are related to $t$, $r$, $f$, $h$ in the coordinate (\ref{SSS}), as
\be
\rd v=\rd t +\frac{\rd r}{\sqrt{fh}}\,,\qquad 
B(r)=\sqrt{\frac{f}{h}}\,.
\ee
For the SSS backgrounds (\ref{SSS}) and (\ref{SSS2}), we can choose the Aether-field configurations, respectively, as
\ba
u^{\mu}\partial_{\mu}
&=& a(r) \partial_t+b(r)\partial_r \\
&=&-\alpha(r)\partial_v-\beta(r) \partial_r\,,
\ea
where $a$, $b$, $\alpha$, and $\beta$ are $r$-dependent functions related to each other as 
\be
a+\frac{b}{\sqrt{fh}}=-\alpha\,,\qquad
b=-\beta\,.
\label{ab}
\ee
From the unit vector constraint (\ref{be1}), we obtain
\be
b^2=(a^2 f -1)h\,, \qquad 1=f\alpha^2-2B\alpha\beta\,,
\label{bso}
\ee
so that the inequality $(a^2 f -1)h \geq 0$ must hold. 
From Eqs.~(\ref{ab}) and (\ref{bso}), we find the following relations
\be
a=-\frac{1+f\alpha^2}{2f \alpha}\,,\qquad 
b=\sqrt{fh} \frac{1-f \alpha^2}{2f\alpha}\,.
\label{ab2}
\ee

On the metric horizon $r_{\rm g}$ satisfying $f(r_{\rm g})=0=h(r_{\rm g})$, the Eddington-Finkelstein coordinate system is regular and thus $\alpha_0\equiv \alpha(r_{\rm g})$ and $\beta_0\equiv \beta(r_{\rm g})$ should be finite for a regular $u^{\mu}$. Furthermore, due to the second of (\ref{bso}), $\beta_0$ takes a finite and non-zero value $\beta_0=-\sqrt{h_1/f_1}/(2\alpha_0)$, where we perform the expansions $f=\sum_{i=1}f_i (r-r_{\rm g})^i$ and $h=\sum_{i=1}h_i (r-r_{\rm g})^i$ around $r=r_{\rm g}$. This, combined with the first of (\ref{bso}), implies that the product $a^2 fh$ is finite and non-zero at $r=r_{\rm g}$ and hence $a$ diverges as $a \propto 1/\sqrt{fh}$ at $r=r_{\rm g}$. 

At sufficiently large distances where the metric components $f$ and $h$ in Eq.~(\ref{SSS}) approach $1$, the Aether-field configuration approaches that on the Minkowski background, i.e., 
\be
a(r) \to 1\,,\qquad b(r) \to 0\,,\qquad 
{\rm as} \quad r \to \infty\,.
\ee
In the same limit, we have 
\be
\alpha (r) \to -1\,,\qquad \beta(r) \to 0\,,\qquad 
{\rm as} \quad r \to \infty\,,
\ee
where we used the correspondence (\ref{ab}) and assumed without loss of generality that $u^{\mu}$ is future-directed 
(the action (\ref{action}) is invariant under the change $u^{\mu} \to -u^{\mu}$).

For the coordinate (\ref{SSS2}), the background Aether field $u_{\mu}$ has non--vanishing components $u_{v}=(1+f\alpha^2)/(2\alpha)$ and $u_{r}=-\alpha \sqrt{f/h}$. 
On defining $\tau$ by $\rd \tau=\rd v+(u_r/u_v) \rd r$, $u^{\mu}$ is orthogonal to constant-$\tau$ hypersurfaces. These hypersurfaces are spacelike due to the timelike property of $u^{\mu}$.
Similarly, one can introduce another coordinate as $\rd \rho=-\rd v-(s_r/s_v) \rd r$, where $s_v=(1-f\alpha^2)/(2\alpha)$ and $s_r=\alpha \sqrt{f/h}$ are non-vanishing components of a unit vector field $s_{\mu}$ orthogonal to $u^{\mu}$. The two variables $\tau$ and $\rho$ are related to the $t$ and $r$ variables in the coordinate system (\ref{SSS}), as\footnote{Instead of $\phi$ and $\psi$ used in Ref.~\cite{Mukohyama:2024vsn}, we adopt the notations $\tau$ and $\rho$ for them to avoid the confusion with the two-dimensional spherical coordinates $\theta$ and $\varphi$.} 
\ba
\rd \tau &=& \rd t +\frac{1}{\sqrt{fh}} 
\frac{1-f \alpha^2}{1+f \alpha^2} \rd r\,,
\label{phi} \\
\rd \rho &=& - \rd t -\frac{1}{\sqrt{fh}} 
\frac{1+f \alpha^2}{1-f \alpha^2} \rd r\,.
\label{psi}
\ea
This shows that the areal distance $r$ depends on the combination $\tau+\rho$, i.e., 
\be
r=r(\tau+\rho)\,.
\ee
On using these relations, the line element (\ref{SSS}) can be converted to the form  
\be
\rd s^{2} =
-f_A^2(r) \rd \tau^2 
+h_A^2 (r) \rd \rho^2 
+r^2 \left( \rd \theta^2+\sin^2 \theta\,
\rd\varphi^2 \right),
\label{phipsi}
\ee
where 
\be
f_A(r) := -\frac{1+f\alpha^2}{2\alpha}\,,\qquad
h_A(r) := -\frac{1-f\alpha^2}{2\alpha}\,.
\label{fhA}
\ee
Then, the two metric components $g_{\tau \tau}=-f_A^2$ and $g_{\rho \rho}=h_A^2$ are always in the regions $g_{\tau \tau} \le 0$ and $g_{\rho \rho} \ge 0$.
This coordinate system corresponds to the Aether-orthogonal frame in which $\tau$ plays the role of the time measured by observers comoving with $u_{\mu}$ (but with the 
non-trivial lapse $f_A(r)$). 
The timelike Aether field is orthogonal to spacelike hypersurfaces characterized by constant values of $\tau$. 
We note that $\tau$ is known as a Khoronon related to the background Aether field, as 
\be
u_{\mu}=-f_A (r)
\partial_{\mu} \tau\,.
\label{uback}
\ee
In the $(t,r)$ coordinate system, this relation can be easily confirmed by using $u_t=-fa$ and $u_r=b/h$ with Eqs.~(\ref{ab2}) and (\ref{phi}).
In the Aether-orthogonal frame, the Aether field has only the $\tau$ component $u_{\tau}=-f_A(r)$. 

On the metric horizon $r=r_{\rm g}$ at which $f \to 0$ with a finite value of $\alpha$, we have $1+f\alpha^2 \to 1$ and $1-f\alpha^2 \to 1$.
At large distances characterized by the asymptotic behavior $f \to 1$ and $\alpha \to -1$, it follows that $1+f \alpha^2 \to 2$ and $1-f \alpha^2 \to +0$. 
We are interested in the case where positive signs of both $1+f\alpha^2$ and $1-f \alpha^2$ are maintained throughout the metric horizon exterior.
In other words, the signs of $f_A$ and $h_A$ are the same. Without loss of generality, we assume that 
\be
f_A>0\,,\qquad h_A>0\,,
\label{sign}
\ee
in which case $\alpha<0$ in Eq.~(\ref{fhA}).

We note that the sign of $1-f \alpha^2$ is positive inside the metric horizon ($f<0$), while there exists a radius $r_{\rm UH}$ at which $1+f \alpha^2$ vanishes, i.e., 
\be
\left( 1+f \alpha^2 \right)|_{r=r_{\rm UH}}=0\,.
\ee
The position $r=r_{\rm UH}$, which is known as the universal horizon, corresponds to the causal boundary of any speed of propagation. 
The universal horizon exists inside the metric horizon, i.e., $r_{\rm UH}<r_{\rm g}$.
For distances $r>r_{\rm UH}$, both $1+f\alpha^2$ and $1-f\alpha^2$ remain positive, so that $g_{\tau \tau}<0$ and $g_{\rho \rho}>0$ in the line element (\ref{phipsi}). 
This means that the linear stability analysis of BHs in the Aether-orthogonal frame is valid throughout the exterior of the universal horizon ($r>r_{\rm UH}$). 
On the other hand, if we use the coordinate system (\ref{SSS}), the two metric components $f$ and $h$ flip signs in the region $r_{\rm UH}<r<r_{\rm g}$ in comparison to those outside the metric horizon. 
Then, the spacelike property of constant $t$ hypersurfaces is not always ensured by choosing the Killing time $t$ and the areal radius $r$ to study the BH stability outside the 
universal horizon \cite{Mukohyama:2024vsn}. 
This problem does not appear in the Aether-orthogonal frame where the timelike Aether field is orthogonal to constant $\tau$ spacelike hypersurfaces.

\section{Black hole perturbations and gauge issues} 
\label{stasec}

We now formulate the theory of  BH perturbations in the background Aether-orthogonal frame given by the line element (\ref{phipsi}). The background spherical symmetry allows us to consider the $m=0$ component of spherical harmonics $Y_{lm}(\theta,\varphi)$ without loss of generality.
In the following, we focus on the mode $m=0$ and use the notation $Y_l(\theta) := Y_{l0} (\theta)$.
The odd-parity perturbations possess the parity $(-1)^{l+1}$, whereas the even-parity perturbations have the parity $(-1)^{l}$ \cite{Regge:1957td, Zerilli:1970se, DeFelice:2011ka, Kobayashi:2012kh, Kobayashi:2014wsa, Kase:2021mix}. 
On the SSS background (\ref{phipsi}), the components of metric perturbations are given by 
\ba
\hspace{-0.5cm}
& &
h_{\tau \tau}=f_A^2 (r) 
H_0 (\tau,\rho) Y_{l} (\theta)\,,\\
\hspace{-0.5cm}
& &
h_{\tau \rho}=h_{\rho \tau}=H_1 (\tau,\rho) 
Y_{l} (\theta)\,,\\
\hspace{-0.5cm}
& &
h_{\tau \theta}=h_{\theta \tau}=h_0 (\tau,\rho)
Y_{l,\theta}(\theta)\,,\\
\hspace{-0.5cm}
& &
h_{\tau \varphi}=h_{\varphi \tau}=-Q (\tau,\rho) 
(\sin \theta)Y_{l,\theta}(\theta)\,,\label{Q}\\
\hspace{-0.5cm}
& &
h_{\rho \rho}=h_A^2(r) H_2(\tau,\rho)Y_{l}(\theta)\,,\\
\hspace{-0.5cm}
& &
h_{\rho \theta}=h_{\theta \rho}=h_1 (\tau,\rho) 
Y_{l,\theta}(\theta)\,,\\
\hspace{-0.5cm}
& &
h_{\rho \varphi}=h_{\varphi \rho}=-W (\tau,\rho) 
(\sin \theta)Y_{l,\theta}(\theta)\,,\label{W}\\
\hspace{-0.5cm}
& &
h_{\theta \theta}=r^2 K(\tau,\rho)Y_{l} (\theta)
+r^2 G(\tau,\rho)Y_{l,\theta \theta} (\theta)\,,
\\
\hspace{-0.5cm}
& &
h_{\varphi \varphi}=r^2 K(\tau,\rho)(\sin^2 \theta) 
Y_{l} (\theta) \nonumber\\
\hspace{-0.5cm}& & \quad \quad \quad
+r^2 G(\tau,\rho)
(\sin \theta)(\cos \theta)Y_{l,\theta} (\theta)\,,
\\
\hspace{-0.5cm}& & 
h_{\theta \varphi}=\frac{1}{2}U(\tau,\rho) 
\left[ (\cos \theta)Y_{l,\theta}(\theta)
-(\sin \theta)Y_{l,\theta \theta} (\theta) 
\right]\,,
\ea
where the summation of $Y_{l}(\theta)$ with respect to the multiples $l$ as well as the subscript ``$l$'' in $H_0$, $H_1$ etc. are omitted, and the notations $Y_{l,\theta}=\rd Y_l(\theta)/\rd \theta$, $Y_{l,\theta \theta}=\rd^2 Y_l(\theta)/\rd \theta^2$ are used. The three fields $Q$, $W$, and $U$, which depend on $\tau$ and $\rho$, are the perturbations in the odd-parity sector, whereas the other seven fields $H_0$, $H_1$, $h_0$, $H_2$, $h_1$, $K$, $G$, which are also functions of $\tau$ and $\rho$, are the perturbations in the even-parity sector. 
The Aether field has the following perturbed components \cite{DeFelice:2023rra, Kase:2023kvq, Chen:2024hkm}
\ba
& &
\delta u_{\tau}= \delta u_0 (\tau,\rho) Y_{l}(\theta)\,,\label{delu0}\\
& &
\delta u_{\rho}= \delta u_1 (\tau,\rho) Y_{l}(\theta)\,,\\
& &
\delta u_{\theta}= \delta u_2 (\tau,\rho) 
Y_{l,\theta}(\theta)\,,\\ 
& &
\delta u_{\varphi}= -\delta u (\tau,\rho)
(\sin \theta)Y_{l,\theta}(\theta)\,,
\label{delu}
\ea
with the background value (\ref{uback}), and $\delta u_0$, $\delta u_1$, $\delta u_2$, $\delta u$ are functions of $\tau$ and $\rho$.
Note that only the component $\delta u_{\varphi}$ belongs 
to the perturbation in the odd-parity sector.

Since the background metric components $f_A^2$, $h_A^2$, and $r^2$ only depend on the combination $r=r(\tau+\rho)$, we have the following properties
\be
f_{A,\tau}=f_{A,\rho}\,,\qquad 
h_{A,\tau}=h_{A,\rho}\,,\qquad 
r_{,\tau}=r_{,\rho}\,,
\ee
where the notations ${\cal F}_{,\tau}:=\partial {\cal F}/\partial \tau$ and ${\cal F}_{,\rho}:=\partial {\cal F}/\partial \rho$ are used for any $r$-dependent quantity. 
In the following, we replace the $\tau$ derivatives of ${\cal F}$ with the $\rho$ derivatives of ${\cal F}$.
As we mentioned in Eq.~(\ref{sign}), we consider the case of positive signs of $f_A$ and $h_A$ without loss of generality.

Since there are gauge DOFs that can be removed from some of the perturbed fields, we consider the following infinitesimal gauge transformation
\be
\tilde{x}_{\mu}=x_{\mu}+\xi_{\mu}\,,
\label{gatrans}
\ee
where the covariant components of $\xi_{\mu}$ are given by 
\ba
\xi_{\tau} &=& {\cal T}(\tau,\rho)Y_{l} (\theta)\,,\\
\xi_{\rho} &=& {\cal R}(\tau,\rho)Y_{l} (\theta)\,,\\
\xi_{\theta} &=&
\Theta (\tau,\rho) Y_{l,\theta} (\theta)\,,
\\
\xi_{\varphi} &=&
-\Lambda (\tau,\rho) (\sin \theta) 
Y_{l,\theta} (\theta)\,,
\ea
with ${\cal T}$, ${\cal R}$, $\Theta$, $\Lambda$ being functions of $\tau$ and $\rho$.
At linear order in the gauge transformation vector $\xi_{\mu}$, metric perturbations in the odd-parity sector transform as 
\ba
\tilde{Q} &=& 
Q-\Lambda_{,\tau}
+\frac{2r_{,\rho}}{r}\Lambda\,,
\label{Qtra}\\
\tilde{W} &=& 
W-\Lambda_{,\rho}
+\frac{2r_{,\rho}}{r}\Lambda\,,
\label{Wtra}\\
\tilde{U} &=&
U-2\Lambda\,.
\ea
Similarly, the transformation law for even-parity metric perturbations is given by 
\ba
\hspace{-0.7cm}
\tilde{H}_0 &=& 
H_0-\frac{2}{f_A^2} \left( {\cal T}_{,\tau}
-\frac{f_{A,\rho}}{f_A} {\cal T}
-\frac{f_A f_{A,\rho}}{h_A^2}{\cal R} \right),\\
\hspace{-0.7cm}
\tilde{H}_1 &=& 
H_1-{\cal T}_{,\rho}+\frac{2f_{A,\rho}}{f_A}{\cal T}
-{\cal R}_{,\tau}+\frac{2h_{A,\rho}}{h_A}{\cal R}
\,,\\
\hspace{-0.7cm}
\tilde{H}_2 &=& H_2-\frac{2}{h_A^2} \left(
{\cal R}_{,\rho}-\frac{h_{A,\rho}}{h_A} 
{\cal R}-\frac{h_A h_{A,\rho}}{f_A^2}{\cal T} \right)
\,,\\
\hspace{-0.7cm}
\tilde{h}_0 &=& h_0-{\cal T}-\Theta_{,\tau}
+\frac{2r_{,\rho}}{r} \Theta\,,\\
\hspace{-0.7cm}
\tilde{h}_1 &=& h_1-{\cal R}-\Theta_{,\rho}
+\frac{2r_{,\rho}}{r} \Theta\,,\\
\hspace{-0.7cm}
\tilde{K} &=& K+\frac{2r_{,\rho}}{rf_A^2}
{\cal T}-\frac{2r_{,\rho}}{rh_A^2}{\cal R}\,,\\
\hspace{-0.7cm}
\tilde{G} &=& G-\frac{2\Theta}{r^2}\,.
\ea

The Aether perturbation $\delta u_{\mu}$ is subject to the following transformation 
\be
\widetilde{\delta u}_{\mu}=\delta u_{\mu} 
-u_{\alpha}\,{\xi^{\alpha}}_{,\mu}
-u_{\mu,\alpha} \xi^{\alpha}\,,
\label{Leedu}
\ee
where $u_{\alpha}$ is the background value given by Eq.~(\ref{uback}). 
On using Eqs.~(\ref{delu0})-(\ref{delu}), the components $\delta u_0$, $\delta u_1$, $\delta u_2$, and $\delta u$ transform, respectively, as 
\ba
\widetilde{\delta u}_{0} &=& 
\delta u_0+\frac{f_{A,\rho}}{f_A^2}{\cal T} 
-\frac{1}{f_A}{\cal T}_{,\tau}
+\frac{f_{A,\rho}}{h_A^2} {\cal R}\,,\\
\widetilde{\delta u}_{1} &=& 
\delta u_1+\frac{2f_{A,\rho}}{f_A^2}{\cal T} 
-\frac{1}{f_A} {\cal T}_{,\rho}\,,\\
\widetilde{\delta u}_{2} &=& 
\delta u_2-\frac{1}{f_A}{\cal T}\,,\\
\widetilde{\delta u} &=& \delta u\,.
\ea
Hence, $\delta u$ is gauge invariant.

Let us consider the multipole modes $l \geq 2$.
In the odd-parity sector, we choose the gauge 
\be
\tilde{U}=0\,,
\ee
under which $\Lambda$ is fixed to be $\Lambda=U/2$. 
Then, the perturbations $\tilde{Q}$ and $\tilde{W}$ in Eqs.~(\ref{Qtra}) and (\ref{Wtra}) are fixed without any residual gauge DOFs.

In the even-parity sector, there are several ways to fix the gauge conditions. Two of them are given by 
\be
\widetilde{\delta u}_{2}=0\,,\qquad 
\tilde{G}=0\,,\qquad \tilde{h}_1=0\,,
\label{gauge1}
\ee
and 
\be
\widetilde{\delta u}_{2}=0\,,\qquad 
\tilde{G}=0\,,\qquad
\tilde{K}=0\,,
\label{gauge2}
\ee
under which ${\cal T}$ and $\Theta$ are fixed to be ${\cal T}=f_A \delta u_2$ and $\Theta=r^2 G/2$. For the gauge choices (\ref{gauge1}) and (\ref{gauge2}), ${\cal R}$ is fixed to be ${\cal R}=h_1-r^2 G_{,\rho}/2$ and 
${\cal R}=rh_A^2K/(2r_{,\rho})+h_A^2 \delta u_2/f_A$, respectively.
A couple of other possible gauge choices are given by 
\be
\widetilde{\delta u}_{0}=0\,,\qquad 
\widetilde{\delta u}_{2}=0\,, \qquad 
\tilde{G}=0\,,
\label{gauge3}
\ee
and
\be
\tilde{h}_0=0\,,\qquad \tilde{K}=0\,,
\qquad \tilde{G}=0\,.
\label{gauge4}
\ee
The gauge condition (\ref{gauge4}) was used in Refs.~\cite{Kobayashi:2014wsa, Kase:2021mix, Kase:2023kvq} for studying BH stability 
in Horndeski theories and 
Maxwell-Horndeski theories. 
In Maxwell-Horndeski theories, the existence of the $U(1)$ gauge symmetry allows one to choose $\widetilde{\delta u}_{2}=0$ besides the gauge conditions (\ref{gauge4}).

For theories with broken $U(1)$ gauge invariance, as in the case of EA gravity, the choice of the gauge condition (\ref{gauge4}) does not allow one to fix $\widetilde{\delta u}_{2}=0$ further. In such cases, if we do not set $\widetilde{\delta u}_{2}=0$ to fix ${\cal T}$, the mixture of a longitudinal scalar mode with transverse vector modes can lead to  difficulty in identifying dynamical DOFs. 
In this sense, it would be preferable to choose either the gauge (\ref{gauge1}) or (\ref{gauge2}) containing the condition $\widetilde{\delta u}_{2}=0$.
In Sec.~\ref{evensec}, we  identify the dynamical DOFs of even-parity perturbations under the gauge choice (\ref{gauge1}).

In the following, we omit the tilde from the gauge-transformed fields and simply write the gauge conditions like $\delta u_2=0$.

\section{Odd-parity perturbations} 
\label{oddsec}

We study the linear stability of BHs against odd-parity perturbations for short-wavelength modes by performing direct calculations in the Aether-orthogonal frame. 
Note that this issue was also addressed in Ref.~\cite{Mukohyama:2024vsn} by using the coordinate transformation properties of dynamical perturbations from (\ref{SSS}) to (\ref{phipsi}).
Indeed, we will show that the conditions for the absence of ghosts and Laplacian instabilities are the same as those 
on the Minkowski background.

By choosing the gauge $U=0$, the odd-parity gravitational perturbations are characterized by $Q(\tau, \rho)$ and $W(\tau, \rho)$ in Eqs.~(\ref{Q}) and (\ref{W}).
The Aether field in the odd-parity sector has a configuration 
\be
u_{\mu} \partial^{\mu}=-f_A(\tau+\rho) 
\partial^{\tau}-\delta u(\tau,\rho) 
(\sin \theta) Y_{l,\theta} (\theta) 
\partial^{\varphi}\,.
\ee
Expanding the action (\ref{action}) up to second order in odd-parity perturbations and performing the integrations with respect to $\theta$ and $\varphi$, the quadratic-order action (up to boundary terms) is given by ${\cal S}^{(2)}_{\rm odd}=(16\pi G_{\ae})^{-1} L \int \rd \tau \rd \rho\,{\cal L}^{(2)}_{\rm odd}$, with the Lagrangian
\begin{widetext}
\ba
{\cal L}^{(2)}_{\rm odd}
&=& 
C_1 \left( W_{,\tau}
-\frac{2r_{,\rho}}{r}W-Q_{,\rho}+\frac{2r_{,\rho}}{r}Q 
\right)^2+2 \left( C_2 \delta u_{,\rho}+C_3 \delta u 
\right)\left( W_{,\tau}
-\frac{2r_{,\rho}}{r}W-Q_{,\rho}+\frac{2r_{,\rho}}{r}Q 
\right)+C_4\delta u_{,\tau}^2+C_5\delta u_{,\rho}^2
\nonumber \\
& &
+(LC_6+\tilde{C}_6) W^2+C_7 W \delta u+C_8 WQ
+(LC_9+\tilde{C}_9) Q^2 
+(LC_{10}+\tilde{C}_{10}) Q\delta u+(L C_{11}+\tilde{C}_{11}) \delta u^2\,,
\label{Lag1}
\ea
\end{widetext}
where 
\be
L := l(l+1)\,,
\ee
and the coefficients $C_i$ and $\tilde{C}_i$ are $r$-dependent functions without containing the $L$ dependence. 
The explicit forms of some of $C_i$'s, which will be used later, are 
\ba
& &
C_1=\frac{1-c_{13}}{2f_A h_A}\,,\quad 
C_2=-\frac{c_{13}}{2h_A}\,,\quad
C_4=\frac{c_{14} h_A}{f_A}\,,\nonumber \\
& &
C_5=-\frac{c_1 f_A}{h_A}\,,\quad
C_6=-\frac{f_A}{2h_A r^2}\,, \quad 
C_9=\frac{(1-c_{13})h_A}{2f_A r^2},\nonumber \\
& &
C_{10}=\frac{c_{13} h_A}{r^2}\,,\quad 
C_{11}=-\frac{c_1 f_A h_A}{r^2}\,,
\ea
where the notation 
\be
c_{ij} := c_i+c_j
\label{cij}
\ee
is used. 
We introduce an auxiliary variable $\chi$ as in the following Lagrangian 
\begin{widetext}
\ba
\tilde{{\cal L}}^{(2)}_{\rm odd}
&=& 
C_1 \left[ 2\chi \left( W_{,\tau}
-\frac{2r_{,\rho}}{r}W-Q_{,\rho}+\frac{2r_{,\rho}}{r}Q 
+\frac{C_2 \delta u_{,\rho}+C_3 \delta u}{C_1}
\right)-\chi^2 \right]
-\frac{(C_2 \delta u_{,\rho}+C_3 \delta u)^2}{C_1}
+C_4\delta u_{,\tau}^2+C_5\delta u_{,\rho}^2
\nonumber \\
& &
+(LC_6+\tilde{C}_6) W^2+C_7 W \delta u+C_8 WQ
+(LC_9+\tilde{C}_9) Q^2 
+(LC_{10}+\tilde{C}_{10}) Q\delta u
+(L C_{11}+\tilde{C}_{11}) \delta u^2\,.
\label{Lag2}
\ea
\end{widetext}
Varying Eq.~(\ref{Lag2}) with respect to $\chi$, it follows that 
\be
\chi=W_{,\tau}
-\frac{2r_{,\rho}}{r}W-Q_{,\rho}+\frac{2r_{,\rho}}{r}Q 
+\frac{C_2 \delta u_{,\rho}+C_3 \delta u}{C_1}\,.
\label{chi}
\ee
Substituting Eq.~(\ref{chi}) into Eq.~(\ref{Lag2}), we find that the Lagrangian (\ref{Lag2}) is equivalent to (\ref{Lag1}). 

We vary Eq.~(\ref{Lag2}) with respect to $W$ and $Q$ and solve their perturbation equations of motion for $W$ and $Q$. 
Then, we can eliminate the terms containing $W$, $Q$, and their $\tau$, $\rho$ derivatives from Eq.~(\ref{Lag2}). 
After  integration by parts, the Lagrangian can be expressed in the form 
\ba
\tilde{{\cal L}}^{(2)}_{\rm odd}
&=&K_{11} \chi_{,\tau}^2
+K_{22} \delta u_{,\tau}^2
+G_{11} \chi_{,\rho}^2
+G_{22} \delta u_{,\rho}^2 \nonumber \\
& &
+M_{11} \chi^2+M_{22}\delta u^2
+2M_{12} \chi \delta u\nonumber \\
& &
+R_{11}\chi_{,\tau}\chi_{,\rho}+P_{12} 
\chi \delta u_{,\tau}+Q_{12} 
\chi \delta u_{,\rho}\,,
\label{Lag3}
\ea
where the coefficients $K_{11}$ etc.\ are $r$-dependent functions.
We can implement further integration by parts, for instance, by antisymmetrizing the terms $P_{12} \chi \delta u_{,\tau}$ and $Q_{12} \chi \delta u_{,\rho}$, but the results will not change.
In the eikonal limit $l \gg 1$, these coefficients reduce to 
\ba
\hspace{-0.8cm}
& &
K_{11}=-\frac{C_1^2}{C_6 L}\,,\qquad 
K_{22}=C_4\,,\\
\hspace{-0.8cm}
& &
G_{11}=-\frac{C_1^2}{C_9 L}\,,\qquad 
G_{22}=C_5-\frac{C_2^2}{C_1}\,,\\
\hspace{-0.8cm}
& &
M_{11}=-C_1\,,\qquad 
M_{22}=\left( C_{11}-\frac{C_{10}^2}{4C_9} 
\right)L\,,\\
\hspace{-0.8cm}
& &
M_{12}=C_3+\frac{C_1[rC_9 C_{10,\rho}
-C_{10}(r C_{9,\rho}+2r_{,\rho}C_9)]}{2r C_9^2},\\
\hspace{-0.8cm}
& &
R_{11}=-\frac{C_1^2 C_8}{C_6 C_9 L^2}\,,\\
\hspace{-0.8cm}
& &
P_{12}=\frac{C_1(C_8 C_{10}-2C_7C_9)}
{2C_6 C_9 L}\,,\\
\hspace{-0.8cm}
& &
Q_{12}=2C_2+\frac{C_1 C_{10}}{C_9}\,.
\ea
From the Lagrangian (\ref{Lag3}), it is clear that there are two degrees of dynamical perturbations, given by $\chi$ and $\delta u$.

The propagation of small-scale perturbations with large angular frequencies $\omega$ and momenta $k$ is known by assuming the solutions to the perturbation equations for $\chi$ and $\delta u$ in the forms 
\be
\chi=\bar{\chi} e^{-i (\omega \tau-k \rho)}\,,\qquad 
\delta u=\bar{\delta u} e^{-i (\omega \tau-k \rho)}\,,
\label{chiu}
\ee
where $\bar{\chi}$ and $\bar{\delta u}$ are constants. The no-ghost conditions for the fields $\chi$ and $\delta u$ are determined by the positivity of the coefficients $K_{11}$ and $K_{22}$, respectively, as
\ba
K_{11} &=& \frac{(1-c_{13})^2 r^2}
{2f_A^3 h_A L}>0\,,
\label{nogoodd1}\\
K_{22} &=& \frac{c_{14}h_A}{f_A}>0\,.
\label{nogoodd2}
\ea
The first inequality (\ref{nogoodd1}) holds for $c_{13} \neq 1$. 
The second equality (\ref{nogoodd2}) is satisfied if
\be
c_{14}>0\,,
\label{nogoodd}
\ee
which coincides with the no-ghost condition of vector perturbations on the Minkowski background \cite{Oost:2018tcv}.

The radial propagation speeds can be determined by considering the modes $\omega r_{\rm g} \approx k r_{\rm g} \gg l \gg 1$. 
In this regime, the dominant contributions to the Lagrangian (\ref{Lag3}) are the first four terms, while the term $R_{11}\chi_{,\tau}\chi_{,\rho}$ is suppressed due to the multipole dependence $R_{11} \propto L^{-2} \propto l^{-4}$. 
The squared radial propagation speed measured in terms of the proper time in the 
Aether-orthogonal frame is defined by
\be
c_{r}^2 = \frac{h_A^2}{f_A^2} \left( 
\frac{\rd \rho}{\rd \tau} \right)^2
=\frac{h_A^2}{f_A^2}\frac{\omega^2}{k^2}\,.
\label{crdef}
\ee
Substituting the solutions (\ref{chiu}) into the perturbation equations for $\chi$ and $\delta u$ arising from the Lagrangian (\ref{Lag3}), we 
obtain the two dispersion relations 
\ba
\omega^2 &=& -\frac{G_{11}}{K_{11}}k^2
=\frac{1}{1-c_{13}}\frac{f_A^2}{h_A^2}k^2\,,\\
\omega^2 &=& -\frac{G_{22}}{K_{22}}k^2
=\frac{2c_1-c_{13}(2c_1-c_{13})}
{2c_{14}(1-c_{13})}\frac{f_A^2}{h_A^2}k^2\,.
\ea
Then, the squared radial propagation speeds of $\chi$ and $\delta u$ are given, respectively, by 
\ba
c_{r1,{\rm odd}}^2 &=& 
\frac{1}{1-c_{13}}\,,\\
c_{r2,{\rm odd}}^2 &=& 
\frac{2c_1-c_{13}(2c_1-c_{13})}
{2c_{14}(1-c_{13})}\,,
\ea
which match those of the tensor and vector perturbations on the Minkowski background, respectively \cite{Jacobson:2004ts, Oost:2018tcv}.

The angular propagation speeds can be derived by considering the modes $l \approx \omega r_{\rm g} \gg k r_{\rm g} \gg 1$. 
In this regime, the two kinetic terms $K_{11} \chi_{,\tau}^2$, $K_{22} \delta u_{,\tau}^2$ and the two mass terms $M_{11} \chi^2$, $M_{22}\delta u^2$ are the dominant contributions to the Lagrangian (\ref{Lag3}), while the other terms including $2M_{12} \chi \delta u$, $P_{12}\chi \delta u_{,\tau}$, $Q_{12} \chi \delta u_{,\rho}$ are suppressed in the limit $l \gg 1$. The leading-order terms defining the equations of motion in the eikonal limit are well understood by performing the field redefinition $\chi=\sqrt{L} \tilde\chi$, 
which ensures that the kinetic matrix elements have the same $L$ dependence 
at leading order.
The squared angular propagation speed measured in terms of the proper time is defined by
\be
c_{\Omega}^2=\frac{r^2}{f_A^2} \left( 
\frac{\rd \theta}{\rd \tau} \right)^2
=\frac{r^2}{f_A^2} \frac{\omega^2}{l^2}\,.
\label{cOdef}
\ee
Then, we have the following two dispersion relations
\ba
\omega^2 &=& -\frac{M_{11}}{K_{11}}
=\frac{1}{1-c_{13}}\frac{f_A^2L}{r^2}\,,\\
\omega^2 &=& -\frac{M_{22}}{K_{22}}
=\frac{2c_1-c_{13}(2c_1-c_{13})}
{2c_{14}(1-c_{13})}\frac{f_A^2L}{r^2}\,.
\ea
Hence the squared angular propagation speeds of $\chi$ and $\delta u$ are given, respectively, by 
\ba
c_{\Omega1,{\rm odd}}^2 &=& 
\frac{1}{1-c_{13}}\,,\\
c_{\Omega2,{\rm odd}}^2 &=& 
\frac{2c_1-c_{13}(2c_1-c_{13})}
{2c_{14}(1-c_{13})}\,,
\ea
which coincide with those of the tensor and vector perturbations on the Minkowski background \cite{Jacobson:2004ts, Oost:2018tcv}. 

Thus, the direct calculation of the second-order action of odd-parity perturbations in the Aether-orthogonal frame leads to the same linear 
stability conditions as those derived in Ref.~\cite{Mukohyama:2024vsn}.

\section{Even-parity perturbations} 
\label{evensec}

We now proceed to study the dynamics of even-parity perturbations. 
From the discussion given in Sec.~\ref{stasec}, one can construct the following gauge-invariant variables
\ba
H_2^{(\rm GI)} &=& H_2-\frac{2}{h_A^2} 
\biggl[ 
\left( \frac{r^2h_{A,\rho}}{2h_A}-r r_{,\rho} 
\right) G_{,\rho}-\frac{r^2}{2}G_{,\rho \rho} \nonumber \\
&&
+h_{1,\rho}-\frac{h_{A,\rho}}{h_A}h_1
-\frac{h_A h_{A,\rho}}{f_A} \delta u_2\biggr]\,,
\label{gain1}\\
\delta u_1^{(\rm GI)} &=& 
\delta u_1+\frac{f_{A,\rho}}{f_A} \delta u_2 
-\delta u_{2,\rho}\,,\\
K^{(\rm GI)} &=& K+\frac{rr_{,\rho}}{h_A^2}G_{,\rho}
-\frac{2r_{,\rho}}{rh_A^2}h_1+\frac{2r_{,\rho}}
{rf_A} \delta u_2\,.
\label{gain3}
\ea
To derive the second-order action of even-parity perturbations, we choose the gauge (\ref{gauge1}), i.e., 
\be
\delta u_2=0\,,\qquad G=0\,,\qquad 
h_1=0\,,
\label{gachoice}
\ee
under which Eqs.~(\ref{gain1})-(\ref{gain3}) reduce, respectively, to $H_2^{(\rm GI)}=H_2$, $\delta u_1^{(\rm GI)}=\delta u_1$, and 
$K^{(\rm GI)}=K$.
In the following, we omit the label ``(GI)'' from the upper subscript. 
The gauge-invariant combinations (\ref{gain1})-(\ref{gain3}) correspond to the 
dynamical perturbations in the tensor, vector, and scalar sectors, respectively.

For the gauge choice (\ref{gachoice}), we have 
\be
g_{\mu \nu} u^{\mu} u^{\nu}
=-1-\frac{Y_l(\theta)}{f_A} 
\left( f_A H_0-2\,\delta u_0 \right)+{\cal O}(\epsilon^2)\,,
\ee
where the terms ${\cal O}(\epsilon^2)$ represent perturbations higher than 
the linear order. From the constraint (\ref{be1}), we obtain
\be
\delta u_0=\frac{1}{2}f_A H_0\,.
\label{u0}
\ee
For the gauge choice (\ref{gachoice}), we expand the action (\ref{action}) up to the second order in even-parity perturbations. After the integration with respect to $\theta$ and $\varphi$ and exploiting the relation (\ref{u0}), we obtain the quadratic-order action containing six perturbed variables $H_0$, $H_1$, $h_0$, $H_2$, $\delta u_1$, $K$ and 
their $\tau, \rho$ derivatives. 
After integration by parts, the second-order action is expressed in the form ${\cal S}^{(2)}_{\rm even}=(16\pi G_{\ae})^{-1} \int \rd \tau \rd \rho\,{\cal L}^{(2)}_{\rm even}$, with the Lagrangian
\be
{\cal L}^{(2)}_{\rm even}=
{\cal L}_{H_0}+{\cal L}_{H_1}+{\cal L}_{h_0}
+{\cal L}_{H_2}+{\cal L}_{\delta u_1}
+{\cal L}_{K}\,,
\label{Leven}
\ee
where 
\begin{widetext}
\ba
{\cal L}_{H_0} &=& a_1 H_{0,\rho}^2
+(La_{2}+a_3) H_0^2+H_{0,\tau} (a_4 K+ 
a_5 H_2)+H_{0,\rho} (a_6 \delta u_{1,\tau}
+a_7 K_{,\rho}+ a_8 H_2+ a_9 \delta u_1 
+a_{10} K+ a_{11} H_1) \nonumber \\
& &
+ L H_0 
(a_{12}K+a_{13} H_2+ a_{14} h_0) 
+ H_0 ( a_{15} K+a_{16} H_1+ a_{17} H_2)\,,
\label{H0Lag} \\
{\cal L}_{H_1} &=& b_1 H_{1,\rho}^2
+(Lb_2+b_3)H_1^2+H_{1,\tau}
(b_4 K_{,\rho}+b_5H_2+b_6 K)
+H_{1,\rho} (b_7 H_{2,\tau}+b_8 K_{,\tau}
+b_9 \delta u_{1,\rho}+b_{10} \delta u_1 \nonumber \\
& &
+ b_{11} K+ b_{12} L h_0) 
+ L H_1 ( b_{13} \delta u_1+b_{14} h_0 ) + H_1 (b_{15}K+b_{16} H_2+ b_{17} \delta u_1)\,,\\
{\cal L}_{h_0} &=& L d_1 h_{0,\rho}^2
+L (L d_2 + d_3) h_0^2 + L d_4 h_{0,\rho} 
\delta u_1 + L h_0 (d_5 H_{2,\tau} 
+ d_6 K_{,\tau} +d_7 \delta u_{1,\rho} 
+ d_8 H_2 + d_9 \delta u_1)\,,\\
{\cal L}_{H_2} &=& e_1 H_{2,\tau}^2
+e_2 H_2^2+H_{2,\tau} (e_3 K_{,\tau}
+e_4 \delta u_{1,\rho}+e_5 K+e_6 \delta u_1)
+e_7 H_{2,\rho}K+e_8 L H_2 K
+ H_2 ( e_9 K+ e_{10} \delta u_1)\,,\\
{\cal L}_{\delta u_1} &=& 
f_1 \delta u_{1,\tau}^2+ f_2 \delta u_{1,\rho}^2 
+ (Lf_3 + f_4) \delta u_1^2 
+ f_5 \delta u_{1,\tau} K+ \delta u_{1,\rho}
(f_6 K_{,\tau} + f_7 K) 
+ \delta u_1(f_8 K_{,\tau} + f_9 K_{,\rho} 
+ f_{10} K)\,,
\label{u1Lag} \\
{\cal L}_{K} &=& g_1 K_{,\tau}^2
+g_2 K_{,\rho}^2\,.
\label{KLag}
\ea
\end{widetext}
The coefficients $a_1, b_1, \cdots$ depend on $\tau+\rho$, without  dependence on $L$. 
Since the fields $H_0$, $H_1$, and $h_0$ do not contain the second-order $\tau$ derivatives such as $h_{0,\tau}^2$, they are regarded as non-dynamical perturbations that can be integrated out from the Lagrangian ${\cal L}^{(2)}_{\rm even}$. 
On the other hand, the perturbations $H_2$, $\delta u_1$, and $K$ are the dynamical fields containing  quadratic-order $\tau$ derivatives such as $H_{2,\tau}^2$.

To study the behavior of short-wavelength perturbations, we assume the solutions to the perturbation equations in the forms 
\ba
& &
H_0=\bar{H}_0 e^{-i (\omega \tau-k \rho)}\,,
\qquad 
H_1=\bar{H}_1 e^{-i (\omega \tau-k \rho)}\,,
\nonumber \\
& &
h_0=\bar{h}_0 e^{-i (\omega \tau-k \rho)}\,,
\qquad 
H_2=\bar{H}_2 e^{-i (\omega \tau-k \rho)}\,,
\nonumber \\
& &
\delta u_1=\bar{\delta u}_1 e^{-i (\omega \tau-k \rho)}\,,
\qquad 
K=\bar{K} e^{-i (\omega \tau-k \rho)}\,,
\label{WKB}
\ea
where $\bar{H}_0$, $\bar{H}_1$, $\bar{h}_0$, $\bar{H}_2$, 
$\bar{\delta u}_1$, and $\bar{K}$ are constants. 
We are interested in the small-scale modes in the range 
\be
\omega^2 r_g^2 \gg 1\,,\qquad
k^2 r_g^2 \gg 1\,,\qquad L \gg 1\,.
\label{smallap}
\ee
By picking up the dominant contributions of each Lagrangian in Eqs.~(\ref{H0Lag})-(\ref{KLag}), we have
\ba
\hspace{-1cm}
{\cal L}_{H_0} &\simeq& 
a_1 H_{0,\rho}^2
+L a_2 H_0^2
+H_{0,\rho} (a_6 \delta u_{1,\tau}
+a_7 K_{,\rho}) \label{LH0ap} \nonumber\\
\hspace{-1cm}
& &
+L H_0 (a_{12}K+a_{13} H_2)\,,\label{LH0}\\
\hspace{-1cm}
{\cal L}_{H_1} &\simeq& 
b_1 H_{1,\rho}^2+Lb_2 H_1^2
+b_4 H_{1,\tau}K_{,\rho}
+H_{1,\rho} (b_7 H_{2,\tau} \nonumber\\
\hspace{-1cm}
& &
+b_8 K_{,\tau}
+b_9 \delta u_{1,\rho}+b_{12}Lh_0) 
+ Lb_{13} H_1 \delta u_1\,,\\
\hspace{-1cm}
{\cal L}_{h_0} &\simeq& L [d_1 h_{0,\rho}^2
+L d_2 h_0^2+d_4 h_{0,\rho} \delta u_1 \nonumber\\
\hspace{-1cm}
& & +h_0 ( d_5 H_{2,\tau}+d_6 K_{,\tau}
+d_7 \delta u_{1,\rho}) ]\,,\label{Lh0} \\
\hspace{-1cm}
{\cal L}_{H_2} &\simeq& e_1 H_{2,\tau}^2
+H_{2,\tau} (e_3 K_{,\tau}+e_4 \delta u_{1,\rho})
+e_8 L H_2 K,\\
\hspace{-1cm}
{\cal L}_{\delta u_1} &\simeq& 
f_1 \delta u_{1,\tau}^2+ f_2 \delta u_{1,\rho}^2 
+Lf_3 \delta u_1^2+f_6 \delta u_{1,\rho}K_{,\tau}\,,
\label{Lu1}\\
\hspace{-1cm}
{\cal L}_{K} &=& g_1 K_{,\tau}^2+g_2 K_{,\rho}^2\,.
\label{LK}
\ea
The Lagrangian ${\cal L}_{H_1}$ contains the term $b_{12}H_{1,\rho}L h_0$, which is of order $L$ for $h_0 \propto 1/k$.
Hence we have not ignored the last four terms in Eq.~(\ref{Lh0}). 
In Appendix A, we show explicit forms of the coefficients in Eqs.~(\ref{LH0})-(\ref{LK}).

In Eqs.~(\ref{LH0ap})-(\ref{LK}), we perform the following integration by parts 
\ba
& &
A \psi_{i, \tau}\psi_{j, \rho}=\frac{1}{2}A
\left( \psi_{i, \tau}\psi_{j, \rho}
+\psi_{j, \tau}\psi_{i, \rho} \right)+\cdots\,,\\
& &
A \psi_{i, \tau}\psi_{j}=\frac{1}{2}A
\left( \psi_{i, \tau}\psi_{j}
-\psi_{j, \tau}\psi_{i} \right)+\cdots\,,\\
& &
A \psi_{i, \rho}\psi_{j}=\frac{1}{2}A
\left( \psi_{i, \rho}\psi_{j}
-\psi_{j, \rho}\psi_{i} \right)+\cdots\,,
\ea
where $A$ depends on $\tau+\rho$, and the dots represent the subleading terms in  the above approximation scheme. 
This allows one to obtain the Lagrangian in a Hermitian form \cite{DeFelice:2023kpl}.
The perturbation equations of motion for the six variables $\psi_i=(H_0, H_1, h_0, H_2, \delta u_1, K)$ follow by varying the total Lagrangian (\ref{Leven}) with Eqs.~(\ref{LH0ap})-(\ref{LK}). 
In doing so, we use an approximation that the coefficients $a_1, b_1, \cdots$ are constants and substitute the WKB-form solutions (\ref{WKB}) into the six perturbation equations of motion.
We solve the first three equations for the fields $\bar{H}_0$, $\bar{H}_1$, and 
$\bar{h}_0$ and then substitute them into the last three equations. 
Then, the field equations for the three dynamical variables (represented by the corresponding constants), 
\be
\vec{\Psi}=\left( \bar{H}_2, \bar{\delta u}_1, 
\bar{K} \right)\,,
\ee
are expressed in the form 
\be
{\bm B}_{\rm even}\vec{\Psi}^{\rm T}={\bm 0}\,,
\label{Beven}
\ee
where ${\bm B}_{\rm even}$ is a $3 \times 3$ Hermitian matrix. 

The no-ghost conditions follow by selecting  terms proportional to $\omega^2$ in ${\bm B}_{\rm even}$. 
Expressing the matrix containing these components as $\omega^2 {\bm B}$, the determinants of the submatrices of ${\bm B}$ must be positive to avoid ghosts. 
Since we have 
\be
B_{12}=B_{21}=0\,,\qquad 
B_{23}=B_{23}=0\,,
\ee
the perturbation $\bar{\delta u}_1$ is decoupled from the other two dynamical perturbations. 
Then, the no-ghost condition for the vector-field perturbation is given by 
\be
{\cal K}_1 := B_{22}
=\frac{2c_{14} r^2 h_A L}
{(k^2 r^2+L h_A^2)f_A}>0\,.
\ee
The other two no-ghost conditions, which correspond to those of tensor and scalar perturbations, are 
\ba
\hspace{-0.9cm}
{\cal K}_2 &:=& B_{11}B_{33}-B_{13}B_{31}
\nonumber \\
\hspace{-0.9cm}
&=& \frac{(2+c_{13} + 3c_2)(1-c_{13})^2
r^4 h_A^6 L^2}{4c_{123}(k^2 r^2+L h_A^2)^2f_A^2}
>0\,,
\ea
and 
\ba
\hspace{-0.9cm}
{\cal K}_3 &:=& {\rm det}\, {\bm B}
={\cal K}_1 {\cal K}_2>0\,.
\ea
These inequalities are satisfied if
\ba
& &
c_{14}>0\,,\label{nogoeven1}\\
& &
\frac{2+c_{13} + 3c_2}{c_{123}}>0\,.
\label{nogoeven2}
\ea
Note that the first condition (\ref{nogoeven1}) is equivalent to the no-ghost condition (\ref{nogoodd}) of the vector-field perturbation in the 
odd-parity sector.

The solutions to Eq.~(\ref{Beven}) with non-vanishing components of $\vec{\Psi}$ are present if 
\be
{\rm det}\,{\bm B}_{\rm even}=0\,.
\ee
This gives the following three dispersion relations 
\ba
\omega^2 &=& c_T^2 \frac{f_A^2 (k^2 r^2+Lh_A^2)}
{h_A^2 r^2}\,,
\label{dis1}\\
\omega^2 &=& c_V^2 \frac{f_A^2 (k^2 r^2+Lh_A^2)}
{h_A^2 r^2}\,,\\
\omega^2 &=& c_S^2 \frac{f_A^2 (k^2 r^2+Lh_A^2)}
{h_A^2 r^2}\,,
\label{dis3}
\ea
where 
\ba
c_T^2 &=&\frac{1}{1-c_{13}}\,,
\label{cT}\\
c_V^2 &=& 
\frac{2c_1-c_{13}(2c_1-c_{13})}
{2c_{14}(1-c_{13})}\,,
\label{cV}\\
c_S^2 &=& \frac{c_{123}(2-c_{14})}{c_{14}(1-c_{13})
(2+c_{13}+3c_2)}\,.
\label{cS}
\ea
The squared radial propagation speeds, which are defined by Eq.~(\ref{crdef}), are 
derived by taking the limit 
$k^2 r^2 \gg L h_A^2$ 
in Eqs.~(\ref{dis1})-(\ref{dis3}). 
Then, we obtain the following three squared propagation speeds
\be
c_{r1}^2=c_T^2\,,\qquad
c_{r2}^2=c_V^2\,,\qquad
c_{r3}^2=c_S^2\,,
\ee
which correspond to those of tensor, vector, and scalar perturbations, respectively.

The squared angular propagation speeds, which are defined by Eq.~(\ref{cOdef}), are known under the limit $L h_A^2 \gg k^2 r^2$ in Eqs.~(\ref{dis1})-(\ref{dis3}). 
Then, the corresponding three squared propagation speeds are
\be
c_{\Omega 1}^2=c_T^2\,,\qquad
c_{\Omega 2}^2=c_V^2\,,\qquad
c_{\Omega 3}^2=c_S^2\,,
\ee
which are equivalent to $c_{r1}^2$, $c_{r2}^2$, and $c_{r3}^2$, respectively. 
These values are the same as those on the squared propagation speeds of tensor, vector, and scalar perturbations on the Minkowski 
background \cite{Jacobson:2004ts,Oost:2018tcv}.
Thus, the perturbative analysis in the Aether-orthogonal frame does not provide new small-scale stability conditions to those known 
in the literature.

The Laplacian instability of even-parity perturbations is absent for
$c_T^2>0$, $c_V^2>0$, and $c_S^2>0$. 
These inequalities as well as the no-ghost conditions (\ref{nogoeven1}) and (\ref{nogoeven2}) are satisfied if 
\ba
& &
0<c_{14}<2\,,
\label{con1}\\
& &
c_{13}<1\,,\\
& &
2c_1-c_{13}(2c_1-c_{13})>0\,,\\
& &
\frac{2+c_{13} + 3c_2}{c_{123}}>0\,.
\label{con4}
\ea
Under these conditions, the linear stability of odd-parity perturbations with short wavelengths is also ensured.

\section{Conclusions}
\label{concludesec}

The EA theory is distinguished from other vector-tensor theories like generalized Proca theories \cite{Tasinato:2014eka, Heisenberg:2014rta} in that there is always a preferred timelike direction along the unit Aether field. 
In the former case, the coordinates of the Killing time $t$ and the areal radius $r$ are not necessarily appropriate for studying the BH stability against linear perturbations.  
This is attributed to the fact that the constant $t$ hypersurfaces are not always spacelike, especially in the region between the universal  and  metric horizons. 

To overcome this problem, we introduced the Aether-orthogonal frame in which the timelike coordinate $\tau$ is related to the Aether field $u_{\mu}$ as $u_{\mu}=-f_A(r)\partial_{\mu}\tau$. Introducing the spatial coordinates $(\rho,\theta,\varphi)$ as well, the line element of the Aether-orthogonal frame is given by Eq.~(\ref{phipsi}), whose $\tau \tau$ and $\rho \rho$ metric components are always in the ranges $g_{\tau \tau}=-f_A^2(r) \leq 0$ and $g_{\rho \rho}=h_A^2 (r) \geq 0$. 
We adopted this line element to study the linear stability of BHs on  SSS backgrounds. 

In Sec.~\ref{stasec}, we considered the general formulation of BH perturbations for both odd- and even-parity perturbations in the Aether-orthogonal frame. We also clarified the transformation properties of all perturbed fields under the infinitesimal coordinate shifts (\ref{gatrans}). 
We discussed possible choices of gauges to fix the four components of $\xi_{\mu}$. 
To compute the second-order perturbed action, we chose the gauge  
$\tilde{U}=0$ for odd-parity modes and $\widetilde{\delta u}_{2}=0$,  
$\tilde{G}=0$, and $\tilde{h}_1=0$ for even-parity modes. 
This completely fixes the gauge DOFs for the multipoles $l \geq 2$.

In Sec.~\ref{oddsec}, we computed the quadratic-order action of odd-parity perturbations on SSS backgrounds in the Aether-orthogonal frame without transforming perturbed fields from the $(t,r)$ coordinates
to the $(\tau,\rho)$ ones (where the latter was performed in Ref.~\cite{Mukohyama:2024vsn}). 
After integrating out non-dynamical variables, there are two dynamical perturbations $\chi$ and $\delta u$ left in the second-order action, where $\chi$ corresponds to a tensor DOF defined by Eq.~(\ref{chi}) and $\delta u$ is a vector DOF. 
We showed that, in the small-scale limit, the propagation speeds as well as no-ghost conditions of $\chi$ and $\delta u$ coincide with those of tensor and vector perturbations on the Minkowski background, respectively.

In Sec.~\ref{evensec}, we studied the propagation of even-parity perturbations by choosing the gauge condition (\ref{gachoice}). Since the unit vector constraint gives the relation $\delta u_0=f_A H_0/2$, there are only 
six perturbation variables that are appearing in the second-order Lagrangian 
${\cal L}^{(2)}_{\rm even}$, see Eq.~(\ref{Leven}) with Eqs.~(\ref{H0Lag})-(\ref{KLag}).
Among them, the three fields $H_2$, $\delta u_1$, and $K$, which contain the quadratic-order $\tau$ derivatives in ${\cal L}^{(2)}_{\rm even}$, are the dynamical perturbations in the tensor, vector, and scalar sectors, respectively. For short-wavelength perturbations satisfying the conditions (\ref{smallap}), the dominant contributions to the second-order Lagrangian are given by Eqs.~(\ref{LH0ap})-(\ref{LK}). 
Substituting the WKB-form solutions (\ref{WKB}) into the perturbation equations and eliminating the non-dynamical variables $\bar{H}_0$, $\bar{H}_1$, and $\bar{h}_0$, the three dynamical variables $\vec{\Psi}=(\bar{H}_2,\bar{\delta u}_1,\bar{K})$ satisfy the algebraic equations of the form (\ref{Beven}). 
The no-ghost conditions and speeds of propagation extracted from the matrix ${\bm B}_{\rm even}$ are the same as those of tensor, vector, and scalar perturbations on the Minkowski background.

In summary, for non-vanishing coupling constants $c_{1,2,3,4}$, we showed that there are in total five propagating DOFs, i.e., tensor and vector DOFs in the odd-parity sector and tensor, vector, and scalar DOFs in the even-parity sector. 
The squared tensor, vector, and scalar propagation speeds are given, respectively, by Eqs.~(\ref{cT}), (\ref{cV}), and (\ref{cS}), with the no-ghost conditions (\ref{nogoeven1}) and (\ref{nogoeven2}). 
The linear stability of BHs is ensured under the four conditions (\ref{con1})-(\ref{con4}).

Since we now have all the linear perturbation equations of motion for both odd- and even-parity sectors in EA theory, it should be straightforward to compute the quasinormal modes of BHs on SSS backgrounds in the Aether-orthogonal frame. 
It will be of interest to investigate whether the reduction of small-scale linear stability conditions to those on the Minkowski background holds or not for more general backgrounds in the presence of a timelike vector field without vorticity.
For this purpose, the geometric optics approximation used in Ref.~\cite{Kubota:2022lbn} may help to understand the behavior of short-wavelength perturbations.
These topics are left for future  works.

\section*{Acknowledgements}

We thank Kei-ichi Maeda for useful discussions.
The work of ADF was supported by the Japan Society for the Promotion of Science Grants-in-Aid for Scientific Research No.~20K03969.
The work of SM was supported in part by the Japan Society for the Promotion of Science (JSPS) Grants-in-Aid for Scientific Research No.~24K07017 and the World Premier International Research Center Initiative (WPI), MEXT, Japan. 
ST was supported by the Grant-in-Aid for Scientific Research Fund of the JSPS No.~22K03642 and Waseda University Special Research Project No.~2024C-474. 
AW was partially supported by a US NSF grant with the grant number: PHY2308845.
CZ was supported by the National Natural Science Foundation of China under Grant No.~12205254 and the Science Foundation of China University of Petroleum, Beijing under Grant No.~2462024BJRC005. 

\section*{Appendix:~Coefficients of the second-order 
Lagrangian of even-parity perturbations}
\renewcommand{\theequation}{A.\arabic{equation}} \setcounter{equation}{0}

The coefficients appearing in Eqs.~(\ref{LH0})-(\ref{LK}) are given by 
\begin{widetext}
\ba
& &
a_1=\frac{c_{14}r^2 f_A}{4h_A}\,,\quad
a_2=\frac{c_{14}f_A h_A}{4}\,,\quad
a_6=-\frac{c_{14}r^2}{h_A}\,,\quad 
a_7=-\frac{r^2 f_A}{h_A}\,,\quad
a_{12}=-\frac{f_A h_A}{2}\,,\quad
a_{13}=a_{12}\,,\nonumber \\
& &
b_1=-\frac{c_{123}r^2}{f_A h_A^3}\,,\quad
b_2=\frac{1-c_{13}}{2f_A h_A}\,,\quad
b_4=\frac{4r^2}{f_A h_A}\,,\quad
b_7=\frac{c_{123}r^2}{f_A h_A}\,,\quad
b_8=\frac{2(c_2-1)r^2}{f_A h_A}\,,\quad
b_9=\frac{2c_{123}r^2}{h_A^3}\,,\nonumber \\
& &
b_{12}=\frac{1+c_{13}+2c_2}{f_A h_A}\,,\quad
b_{13}=\frac{c_{13}}{h_A}\,,\nonumber \\
& &
d_1=\frac{1-c_{13}}{2f_A h_A}\,,\quad
d_2=-\frac{c_{123}h_A}{r^2 f_A}\,,\quad
d_4=\frac{c_{13}}{h_A}\,,\quad
d_5=-\frac{(1+c_2)h_A}{f_A}\,,\quad
d_6=-h_A^2b_{12}\,,\quad
d_7=-\frac{2c_2}{h_A}\,,\nonumber \\
& &
e_1=-\frac{c_{123}r^2h_A}{4f_A}\,,\quad
e_3=-\frac{(1+c_2)r^2 h_A}{f_A}\,,\quad
e_4=-\frac{c_{123}r^2}{h_A}\,,\quad
e_8=\frac{f_A h_A}{2}\,,\nonumber \\
& &
f_1=\frac{c_{14}r^2}{f_A h_A}\,,\quad
f_2=-\frac{c_{123}r^2 f_A}{h_A^3}\,,\quad
f_3=-\frac{c_1f_A}{h_A}\,,\quad
f_6=-\frac{2c_2 r^2}{h_A}\,,\quad
g_1=-\frac{r^2 h_A^2}{2}b_{12}\,,\quad
g_2=\frac{r^2 f_A}{2h_A}\,,
\ea
where we used the notation (\ref{cij}) and 
$c_{ijk}:=c_i+c_j+c_k$.
\end{widetext}

\bibliographystyle{mybibstyle}
\bibliography{bib}

\begin{thebibliography}{69}%
\makeatletter
\providecommand \@ifxundefined [1]{%
 \@ifx{#1\undefined}
}%
\providecommand \@ifnum [1]{%
 \ifnum #1\expandafter \@firstoftwo
 \else \expandafter \@secondoftwo
 \fi
}%
\providecommand \@ifx [1]{%
 \ifx #1\expandafter \@firstoftwo
 \else \expandafter \@secondoftwo
 \fi
}%
\providecommand \natexlab [1]{#1}%
\providecommand \enquote  [1]{``#1''}%
\providecommand \bibnamefont  [1]{#1}%
\providecommand \bibfnamefont [1]{#1}%
\providecommand \citenamefont [1]{#1}%
\providecommand \href@noop [0]{\@secondoftwo}%
\providecommand \href [0]{\begingroup \@sanitize@url \@href}%
\providecommand \@href[1]{\@@startlink{#1}\@@href}%
\providecommand \@@href[1]{\endgroup#1\@@endlink}%
\providecommand \@sanitize@url [0]{\catcode `\\12\catcode `\$12\catcode
  `\&12\catcode `\#12\catcode `\^12\catcode `\_12\catcode `\%12\relax}%
\providecommand \@@startlink[1]{}%
\providecommand \@@endlink[0]{}%
\providecommand \url  [0]{\begingroup\@sanitize@url \@url }%
\providecommand \@url [1]{\endgroup\@href {#1}{\urlprefix }}%
\providecommand \urlprefix  [0]{URL }%
\providecommand \Eprint [0]{\href }%
\providecommand \doibase [0]{http://dx.doi.org/}%
\providecommand \selectlanguage [0]{\@gobble}%
\providecommand \bibinfo  [0]{\@secondoftwo}%
\providecommand \bibfield  [0]{\@secondoftwo}%
\providecommand \translation [1]{[#1]}%
\providecommand \BibitemOpen [0]{}%
\providecommand \bibitemStop [0]{}%
\providecommand \bibitemNoStop [0]{.\EOS\space}%
\providecommand \EOS [0]{\spacefactor3000\relax}%
\providecommand \BibitemShut  [1]{\csname bibitem#1\endcsname}%
\let\auto@bib@innerbib\@empty
\bibitem [{\citenamefont {Chadha}\ and\ \citenamefont
  {Nielsen}(1983)}]{Chadha:1982qq}%
  \BibitemOpen
  \bibfield  {author} {\bibinfo {author} {\bibfnamefont {S.}~\bibnamefont
  {Chadha}} and \bibinfo {author} {\bibfnamefont {H.~B.}\ \bibnamefont
  {Nielsen}},\ }\href {\doibase 10.1016/0550-3213(83)90081-0} {\bibfield
  {journal} {\bibinfo  {journal} {\emph {Nucl. Phys. B}}\ }\textbf {\bibinfo
  {volume} {217}},\ \bibinfo {pages} {125} (\bibinfo {year}
  {1983})}\BibitemShut {NoStop}%
\bibitem [{\citenamefont {Kostelecky}\ and\ \citenamefont
  {Samuel}(1989)}]{Kostelecky:1988zi}%
  \BibitemOpen
  \bibfield  {author} {\bibinfo {author} {\bibfnamefont {V.~A.}\ \bibnamefont
  {Kostelecky}} and \bibinfo {author} {\bibfnamefont {S.}~\bibnamefont
  {Samuel}},\ }\href {\doibase 10.1103/PhysRevD.39.683} {\bibfield  {journal}
  {\bibinfo  {journal} {\emph {Phys. Rev. D}}\ }\textbf {\bibinfo {volume}
  {39}},\ \bibinfo {pages} {683} (\bibinfo {year} {1989})}\BibitemShut
  {NoStop}%
\bibitem [{\citenamefont {Gambini}\ and\ \citenamefont
  {Pullin}(1999)}]{Gambini:1998it}%
  \BibitemOpen
  \bibfield  {author} {\bibinfo {author} {\bibfnamefont {R.}~\bibnamefont
  {Gambini}} and \bibinfo {author} {\bibfnamefont {J.}~\bibnamefont {Pullin}},\
  }\href {\doibase 10.1103/PhysRevD.59.124021} {\bibfield  {journal} {\bibinfo
  {journal} {\emph {Phys. Rev. D}}\ }\textbf {\bibinfo {volume} {59}},\
  \bibinfo {pages} {124021} (\bibinfo {year} {1999})},\ \Eprint
  {http://arxiv.org/abs/gr-qc/9809038} {arXiv:gr-qc/9809038} \BibitemShut
  {NoStop}%
\bibitem [{\citenamefont {Douglas}\ and\ \citenamefont
  {Nekrasov}(2001)}]{Douglas:2001ba}%
  \BibitemOpen
  \bibfield  {author} {\bibinfo {author} {\bibfnamefont {M.~R.}\ \bibnamefont
  {Douglas}} and \bibinfo {author} {\bibfnamefont {N.~A.}\ \bibnamefont
  {Nekrasov}},\ }\href {\doibase 10.1103/RevModPhys.73.977} {\bibfield
  {journal} {\bibinfo  {journal} {\emph {Rev. Mod. Phys.}}\ }\textbf {\bibinfo
  {volume} {73}},\ \bibinfo {pages} {977} (\bibinfo {year} {2001})},\ \Eprint
  {http://arxiv.org/abs/hep-th/0106048} {arXiv:hep-th/0106048} \BibitemShut
  {NoStop}%
\bibitem [{\citenamefont {Carroll}\ \emph {et~al.}(2001)\citenamefont
  {Carroll}, \citenamefont {Harvey}, \citenamefont {Kostelecky}, \citenamefont
  {Lane},\ and\ \citenamefont {Okamoto}}]{Carroll:2001ws}%
  \BibitemOpen
  \bibfield  {author} {\bibinfo {author} {\bibfnamefont {S.~M.}\ \bibnamefont
  {Carroll}}, \bibinfo {author} {\bibfnamefont {J.~A.}\ \bibnamefont {Harvey}},
  \bibinfo {author} {\bibfnamefont {V.~A.}\ \bibnamefont {Kostelecky}},
  \bibinfo {author} {\bibfnamefont {C.~D.}\ \bibnamefont {Lane}},  and \bibinfo
  {author} {\bibfnamefont {T.}~\bibnamefont {Okamoto}},\ }\href {\doibase
  10.1103/PhysRevLett.87.141601} {\bibfield  {journal} {\bibinfo  {journal}
  {\emph {Phys. Rev. Lett.}}\ }\textbf {\bibinfo {volume} {87}},\ \bibinfo
  {pages} {141601} (\bibinfo {year} {2001})},\ \Eprint
  {http://arxiv.org/abs/hep-th/0105082} {arXiv:hep-th/0105082} \BibitemShut
  {NoStop}%
\bibitem [{\citenamefont {Amelino-Camelia}(2013)}]{Amelino-Camelia:2008aez}%
  \BibitemOpen
  \bibfield  {author} {\bibinfo {author} {\bibfnamefont {G.}~\bibnamefont
  {Amelino-Camelia}},\ }\href {\doibase 10.12942/lrr-2013-5} {\bibfield
  {journal} {\bibinfo  {journal} {\emph {Living Rev. Rel.}}\ }\textbf {\bibinfo
  {volume} {16}},\ \bibinfo {pages} {5} (\bibinfo {year} {2013})},\ \Eprint
  {http://arxiv.org/abs/0806.0339} {arXiv:0806.0339 [gr-qc]} \BibitemShut
  {NoStop}%
\bibitem [{\citenamefont {Horava}(2009)}]{Horava:2009uw}%
  \BibitemOpen
  \bibfield  {author} {\bibinfo {author} {\bibfnamefont {P.}~\bibnamefont
  {Horava}},\ }\href {\doibase 10.1103/PhysRevD.79.084008} {\bibfield
  {journal} {\bibinfo  {journal} {\emph {Phys. Rev. D}}\ }\textbf {\bibinfo
  {volume} {79}},\ \bibinfo {pages} {084008} (\bibinfo {year} {2009})},\
  \Eprint {http://arxiv.org/abs/0901.3775} {arXiv:0901.3775 [hep-th]}
  \BibitemShut {NoStop}%
\bibitem [{\citenamefont {Gasperini}(1987)}]{Gasperini:1987nq}%
  \BibitemOpen
  \bibfield  {author} {\bibinfo {author} {\bibfnamefont {M.}~\bibnamefont
  {Gasperini}},\ }\href {\doibase 10.1088/0264-9381/4/2/026} {\bibfield
  {journal} {\bibinfo  {journal} {\emph {Class. Quant. Grav.}}\ }\textbf
  {\bibinfo {volume} {4}},\ \bibinfo {pages} {485} (\bibinfo {year}
  {1987})}\BibitemShut {NoStop}%
\bibitem [{\citenamefont {Jacobson}\ and\ \citenamefont
  {Mattingly}(2001)}]{Jacobson:2000xp}%
  \BibitemOpen
  \bibfield  {author} {\bibinfo {author} {\bibfnamefont {T.}~\bibnamefont
  {Jacobson}} and \bibinfo {author} {\bibfnamefont {D.}~\bibnamefont
  {Mattingly}},\ }\href {\doibase 10.1103/PhysRevD.64.024028} {\bibfield
  {journal} {\bibinfo  {journal} {\emph {Phys. Rev. D}}\ }\textbf {\bibinfo
  {volume} {64}},\ \bibinfo {pages} {024028} (\bibinfo {year} {2001})},\
  \Eprint {http://arxiv.org/abs/gr-qc/0007031} {arXiv:gr-qc/0007031}
  \BibitemShut {NoStop}%
\bibitem [{\citenamefont {Blas}\ \emph {et~al.}(2010)\citenamefont {Blas},
  \citenamefont {Pujolas},\ and\ \citenamefont {Sibiryakov}}]{Blas:2009qj}%
  \BibitemOpen
  \bibfield  {author} {\bibinfo {author} {\bibfnamefont {D.}~\bibnamefont
  {Blas}}, \bibinfo {author} {\bibfnamefont {O.}~\bibnamefont {Pujolas}},  and
  \bibinfo {author} {\bibfnamefont {S.}~\bibnamefont {Sibiryakov}},\ }\href
  {\doibase 10.1103/PhysRevLett.104.181302} {\bibfield  {journal} {\bibinfo
  {journal} {\emph {Phys. Rev. Lett.}}\ }\textbf {\bibinfo {volume} {104}},\
  \bibinfo {pages} {181302} (\bibinfo {year} {2010})},\ \Eprint
  {http://arxiv.org/abs/0909.3525} {arXiv:0909.3525 [hep-th]} \BibitemShut
  {NoStop}%
\bibitem [{\citenamefont {Jacobson}(2010)}]{Jacobson:2010mx}%
  \BibitemOpen
  \bibfield  {author} {\bibinfo {author} {\bibfnamefont {T.}~\bibnamefont
  {Jacobson}},\ }\href {\doibase 10.1103/PhysRevD.81.101502} {\bibfield
  {journal} {\bibinfo  {journal} {\emph {Phys. Rev. D}}\ }\textbf {\bibinfo
  {volume} {81}},\ \bibinfo {pages} {101502} (\bibinfo {year} {2010})},\
  \bibinfo {note} {[Erratum: Phys.Rev.D 82, 129901 (2010)]},\ \Eprint
  {http://arxiv.org/abs/1001.4823} {arXiv:1001.4823 [hep-th]} \BibitemShut
  {NoStop}%
\bibitem [{\citenamefont {Haghani}\ \emph {et~al.}(2014)\citenamefont
  {Haghani}, \citenamefont {Harko}, \citenamefont {Sepangi},\ and\
  \citenamefont {Shahidi}}]{Haghani:2014ita}%
  \BibitemOpen
  \bibfield  {author} {\bibinfo {author} {\bibfnamefont {Z.}~\bibnamefont
  {Haghani}}, \bibinfo {author} {\bibfnamefont {T.}~\bibnamefont {Harko}},
  \bibinfo {author} {\bibfnamefont {H.~R.}\ \bibnamefont {Sepangi}},  and
  \bibinfo {author} {\bibfnamefont {S.}~\bibnamefont {Shahidi}},\ }\Eprint
  {http://arxiv.org/abs/1404.7689} {arXiv:1404.7689 [gr-qc]} \BibitemShut
  {NoStop}%
\bibitem [{\citenamefont {Jacobson}\ and\ \citenamefont
  {Speranza}(2014)}]{Jacobson:2014mda}%
  \BibitemOpen
  \bibfield  {author} {\bibinfo {author} {\bibfnamefont {T.}~\bibnamefont
  {Jacobson}} and \bibinfo {author} {\bibfnamefont {A.~J.}\ \bibnamefont
  {Speranza}},\ }\Eprint {http://arxiv.org/abs/1405.6351} {arXiv:1405.6351
  [gr-qc]} \BibitemShut {NoStop}%
\bibitem [{\citenamefont {Mukohyama}(2010)}]{Mukohyama:2010xz}%
  \BibitemOpen
  \bibfield  {author} {\bibinfo {author} {\bibfnamefont {S.}~\bibnamefont
  {Mukohyama}},\ }\href {\doibase 10.1088/0264-9381/27/22/223101} {\bibfield
  {journal} {\bibinfo  {journal} {\emph {Class. Quant. Grav.}}\ }\textbf
  {\bibinfo {volume} {27}},\ \bibinfo {pages} {223101} (\bibinfo {year}
  {2010})},\ \Eprint {http://arxiv.org/abs/1007.5199} {arXiv:1007.5199
  [hep-th]} \BibitemShut {NoStop}%
\bibitem [{\citenamefont {Afshordi}\ \emph {et~al.}(2007)\citenamefont
  {Afshordi}, \citenamefont {Chung},\ and\ \citenamefont
  {Geshnizjani}}]{Afshordi:2006ad}%
  \BibitemOpen
  \bibfield  {author} {\bibinfo {author} {\bibfnamefont {N.}~\bibnamefont
  {Afshordi}}, \bibinfo {author} {\bibfnamefont {D.~J.~H.}\ \bibnamefont
  {Chung}},  and \bibinfo {author} {\bibfnamefont {G.}~\bibnamefont
  {Geshnizjani}},\ }\href {\doibase 10.1103/PhysRevD.75.083513} {\bibfield
  {journal} {\bibinfo  {journal} {\emph {Phys. Rev. D}}\ }\textbf {\bibinfo
  {volume} {75}},\ \bibinfo {pages} {083513} (\bibinfo {year} {2007})},\
  \Eprint {http://arxiv.org/abs/hep-th/0609150} {arXiv:hep-th/0609150}
  \BibitemShut {NoStop}%
\bibitem [{\citenamefont {Bhattacharyya}\ \emph {et~al.}(2018)\citenamefont
  {Bhattacharyya}, \citenamefont {Coates}, \citenamefont {Colombo},
  \citenamefont {Gumrukcuoglu},\ and\ \citenamefont
  {Sotiriou}}]{Bhattacharyya:2016mah}%
  \BibitemOpen
  \bibfield  {author} {\bibinfo {author} {\bibfnamefont {J.}~\bibnamefont
  {Bhattacharyya}}, \bibinfo {author} {\bibfnamefont {A.}~\bibnamefont
  {Coates}}, \bibinfo {author} {\bibfnamefont {M.}~\bibnamefont {Colombo}},
  \bibinfo {author} {\bibfnamefont {A.~E.}\ \bibnamefont {Gumrukcuoglu}},  and
  \bibinfo {author} {\bibfnamefont {T.~P.}\ \bibnamefont {Sotiriou}},\ }\href
  {\doibase 10.1103/PhysRevD.97.064020} {\bibfield  {journal} {\bibinfo
  {journal} {\emph {Phys. Rev. D}}\ }\textbf {\bibinfo {volume} {97}},\
  \bibinfo {pages} {064020} (\bibinfo {year} {2018})},\ \Eprint
  {http://arxiv.org/abs/1612.01824} {arXiv:1612.01824 [hep-th]} \BibitemShut
  {NoStop}%
\bibitem [{\citenamefont {Gripaios}(2004)}]{Gripaios:2004ms}%
  \BibitemOpen
  \bibfield  {author} {\bibinfo {author} {\bibfnamefont {B.~M.}\ \bibnamefont
  {Gripaios}},\ }\href {\doibase 10.1088/1126-6708/2004/10/069} {\bibfield
  {journal} {\bibinfo  {journal} {\emph {JHEP}}\ }\textbf {\bibinfo {volume}
  {10}},\ \bibinfo {pages} {069} (\bibinfo {year} {2004})},\ \Eprint
  {http://arxiv.org/abs/hep-th/0408127} {arXiv:hep-th/0408127} \BibitemShut
  {NoStop}%
\bibitem [{\citenamefont {Zlosnik}\ \emph {et~al.}(2007)\citenamefont
  {Zlosnik}, \citenamefont {Ferreira},\ and\ \citenamefont
  {Starkman}}]{Zlosnik:2006zu}%
  \BibitemOpen
  \bibfield  {author} {\bibinfo {author} {\bibfnamefont {T.~G.}\ \bibnamefont
  {Zlosnik}}, \bibinfo {author} {\bibfnamefont {P.~G.}\ \bibnamefont
  {Ferreira}},  and \bibinfo {author} {\bibfnamefont {G.~D.}\ \bibnamefont
  {Starkman}},\ }\href {\doibase 10.1103/PhysRevD.75.044017} {\bibfield
  {journal} {\bibinfo  {journal} {\emph {Phys. Rev. D}}\ }\textbf {\bibinfo
  {volume} {75}},\ \bibinfo {pages} {044017} (\bibinfo {year} {2007})},\
  \Eprint {http://arxiv.org/abs/astro-ph/0607411} {arXiv:astro-ph/0607411}
  \BibitemShut {NoStop}%
\bibitem [{\citenamefont {Kanno}\ and\ \citenamefont
  {Soda}(2006)}]{Kanno:2006ty}%
  \BibitemOpen
  \bibfield  {author} {\bibinfo {author} {\bibfnamefont {S.}~\bibnamefont
  {Kanno}} and \bibinfo {author} {\bibfnamefont {J.}~\bibnamefont {Soda}},\
  }\href {\doibase 10.1103/PhysRevD.74.063505} {\bibfield  {journal} {\bibinfo
  {journal} {\emph {Phys. Rev. D}}\ }\textbf {\bibinfo {volume} {74}},\
  \bibinfo {pages} {063505} (\bibinfo {year} {2006})},\ \Eprint
  {http://arxiv.org/abs/hep-th/0604192} {arXiv:hep-th/0604192} \BibitemShut
  {NoStop}%
\bibitem [{\citenamefont {Chesler}\ and\ \citenamefont
  {Loeb}(2017)}]{Chesler:2017khz}%
  \BibitemOpen
  \bibfield  {author} {\bibinfo {author} {\bibfnamefont {P.~M.}\ \bibnamefont
  {Chesler}} and \bibinfo {author} {\bibfnamefont {A.}~\bibnamefont {Loeb}},\
  }\href {\doibase 10.1103/PhysRevLett.119.031102} {\bibfield  {journal}
  {\bibinfo  {journal} {\emph {Phys. Rev. Lett.}}\ }\textbf {\bibinfo {volume}
  {119}},\ \bibinfo {pages} {031102} (\bibinfo {year} {2017})},\ \Eprint
  {http://arxiv.org/abs/1704.05116} {arXiv:1704.05116 [astro-ph.HE]}
  \BibitemShut {NoStop}%
\bibitem [{\citenamefont {Jacobson}\ and\ \citenamefont
  {Mattingly}(2004)}]{Jacobson:2004ts}%
  \BibitemOpen
  \bibfield  {author} {\bibinfo {author} {\bibfnamefont {T.}~\bibnamefont
  {Jacobson}} and \bibinfo {author} {\bibfnamefont {D.}~\bibnamefont
  {Mattingly}},\ }\href {\doibase 10.1103/PhysRevD.70.024003} {\bibfield
  {journal} {\bibinfo  {journal} {\emph {Phys. Rev. D}}\ }\textbf {\bibinfo
  {volume} {70}},\ \bibinfo {pages} {024003} (\bibinfo {year} {2004})},\
  \Eprint {http://arxiv.org/abs/gr-qc/0402005} {arXiv:gr-qc/0402005}
  \BibitemShut {NoStop}%
\bibitem [{\citenamefont {Elliott}\ \emph {et~al.}(2005)\citenamefont
  {Elliott}, \citenamefont {Moore},\ and\ \citenamefont
  {Stoica}}]{Elliott:2005va}%
  \BibitemOpen
  \bibfield  {author} {\bibinfo {author} {\bibfnamefont {J.~W.}\ \bibnamefont
  {Elliott}}, \bibinfo {author} {\bibfnamefont {G.~D.}\ \bibnamefont {Moore}},
  and \bibinfo {author} {\bibfnamefont {H.}~\bibnamefont {Stoica}},\ }\href
  {\doibase 10.1088/1126-6708/2005/08/066} {\bibfield  {journal} {\bibinfo
  {journal} {\emph {JHEP}}\ }\textbf {\bibinfo {volume} {08}},\ \bibinfo
  {pages} {066} (\bibinfo {year} {2005})},\ \Eprint
  {http://arxiv.org/abs/hep-ph/0505211} {arXiv:hep-ph/0505211} \BibitemShut
  {NoStop}%
\bibitem [{\citenamefont {Abbott}\ \emph {et~al.}(2017)\citenamefont {Abbott}
  \emph {et~al.}}]{LIGOScientific:2017zic}%
  \BibitemOpen
  \bibfield  {author} {\bibinfo {author} {\bibfnamefont {B.~P.}\ \bibnamefont
  {Abbott}} \emph {et~al.} (\bibinfo {collaboration} {LIGO Scientific, Virgo,
  Fermi-GBM, INTEGRAL}),\ }\href {\doibase 10.3847/2041-8213/aa920c} {\bibfield
   {journal} {\bibinfo  {journal} {\emph {Astrophys. J. Lett.}}\ }\textbf
  {\bibinfo {volume} {848}},\ \bibinfo {pages} {L13} (\bibinfo {year}
  {2017})},\ \Eprint {http://arxiv.org/abs/1710.05834} {arXiv:1710.05834
  [astro-ph.HE]} \BibitemShut {NoStop}%
\bibitem [{\citenamefont {Gong}\ \emph {et~al.}(2018)\citenamefont {Gong},
  \citenamefont {Hou}, \citenamefont {Liang},\ and\ \citenamefont
  {Papantonopoulos}}]{Gong:2018cgj}%
  \BibitemOpen
  \bibfield  {author} {\bibinfo {author} {\bibfnamefont {Y.}~\bibnamefont
  {Gong}}, \bibinfo {author} {\bibfnamefont {S.}~\bibnamefont {Hou}}, \bibinfo
  {author} {\bibfnamefont {D.}~\bibnamefont {Liang}},  and \bibinfo {author}
  {\bibfnamefont {E.}~\bibnamefont {Papantonopoulos}},\ }\href {\doibase
  10.1103/PhysRevD.97.084040} {\bibfield  {journal} {\bibinfo  {journal} {\emph
  {Phys. Rev. D}}\ }\textbf {\bibinfo {volume} {97}},\ \bibinfo {pages}
  {084040} (\bibinfo {year} {2018})},\ \Eprint
  {http://arxiv.org/abs/1801.03382} {arXiv:1801.03382 [gr-qc]} \BibitemShut
  {NoStop}%
\bibitem [{\citenamefont {Oost}\ \emph {et~al.}(2018)\citenamefont {Oost},
  \citenamefont {Mukohyama},\ and\ \citenamefont {Wang}}]{Oost:2018tcv}%
  \BibitemOpen
  \bibfield  {author} {\bibinfo {author} {\bibfnamefont {J.}~\bibnamefont
  {Oost}}, \bibinfo {author} {\bibfnamefont {S.}~\bibnamefont {Mukohyama}},
  and \bibinfo {author} {\bibfnamefont {A.}~\bibnamefont {Wang}},\ }\href
  {\doibase 10.1103/PhysRevD.97.124023} {\bibfield  {journal} {\bibinfo
  {journal} {\emph {Phys. Rev. D}}\ }\textbf {\bibinfo {volume} {97}},\
  \bibinfo {pages} {124023} (\bibinfo {year} {2018})},\ \Eprint
  {http://arxiv.org/abs/1802.04303} {arXiv:1802.04303 [gr-qc]} \BibitemShut
  {NoStop}%
\bibitem [{\citenamefont {Foster}\ and\ \citenamefont
  {Jacobson}(2006)}]{Foster:2005dk}%
  \BibitemOpen
  \bibfield  {author} {\bibinfo {author} {\bibfnamefont {B.~Z.}\ \bibnamefont
  {Foster}} and \bibinfo {author} {\bibfnamefont {T.}~\bibnamefont
  {Jacobson}},\ }\href {\doibase 10.1103/PhysRevD.73.064015} {\bibfield
  {journal} {\bibinfo  {journal} {\emph {Phys. Rev. D}}\ }\textbf {\bibinfo
  {volume} {73}},\ \bibinfo {pages} {064015} (\bibinfo {year} {2006})},\
  \Eprint {http://arxiv.org/abs/gr-qc/0509083} {arXiv:gr-qc/0509083}
  \BibitemShut {NoStop}%
\bibitem [{\citenamefont {Carroll}\ and\ \citenamefont
  {Lim}(2004)}]{Carroll:2004ai}%
  \BibitemOpen
  \bibfield  {author} {\bibinfo {author} {\bibfnamefont {S.~M.}\ \bibnamefont
  {Carroll}} and \bibinfo {author} {\bibfnamefont {E.~A.}\ \bibnamefont
  {Lim}},\ }\href {\doibase 10.1103/PhysRevD.70.123525} {\bibfield  {journal}
  {\bibinfo  {journal} {\emph {Phys. Rev. D}}\ }\textbf {\bibinfo {volume}
  {70}},\ \bibinfo {pages} {123525} (\bibinfo {year} {2004})},\ \Eprint
  {http://arxiv.org/abs/hep-th/0407149} {arXiv:hep-th/0407149} \BibitemShut
  {NoStop}%
\bibitem [{\citenamefont {Foster}(2007)}]{Foster:2007gr}%
  \BibitemOpen
  \bibfield  {author} {\bibinfo {author} {\bibfnamefont {B.~Z.}\ \bibnamefont
  {Foster}},\ }\href {\doibase 10.1103/PhysRevD.76.084033} {\bibfield
  {journal} {\bibinfo  {journal} {\emph {Phys. Rev. D}}\ }\textbf {\bibinfo
  {volume} {76}},\ \bibinfo {pages} {084033} (\bibinfo {year} {2007})},\
  \Eprint {http://arxiv.org/abs/0706.0704} {arXiv:0706.0704 [gr-qc]}
  \BibitemShut {NoStop}%
\bibitem [{\citenamefont {Yagi}\ \emph
  {et~al.}(2014{\natexlab{a}})\citenamefont {Yagi}, \citenamefont {Blas},
  \citenamefont {Yunes},\ and\ \citenamefont {Barausse}}]{Yagi:2013qpa}%
  \BibitemOpen
  \bibfield  {author} {\bibinfo {author} {\bibfnamefont {K.}~\bibnamefont
  {Yagi}}, \bibinfo {author} {\bibfnamefont {D.}~\bibnamefont {Blas}}, \bibinfo
  {author} {\bibfnamefont {N.}~\bibnamefont {Yunes}},  and \bibinfo {author}
  {\bibfnamefont {E.}~\bibnamefont {Barausse}},\ }\href {\doibase
  10.1103/PhysRevLett.112.161101} {\bibfield  {journal} {\bibinfo  {journal}
  {\emph {Phys. Rev. Lett.}}\ }\textbf {\bibinfo {volume} {112}},\ \bibinfo
  {pages} {161101} (\bibinfo {year} {2014}{\natexlab{a}})},\ \Eprint
  {http://arxiv.org/abs/1307.6219} {arXiv:1307.6219 [gr-qc]} \BibitemShut
  {NoStop}%
\bibitem [{\citenamefont {Yagi}\ \emph
  {et~al.}(2014{\natexlab{b}})\citenamefont {Yagi}, \citenamefont {Blas},
  \citenamefont {Barausse},\ and\ \citenamefont {Yunes}}]{Yagi:2013ava}%
  \BibitemOpen
  \bibfield  {author} {\bibinfo {author} {\bibfnamefont {K.}~\bibnamefont
  {Yagi}}, \bibinfo {author} {\bibfnamefont {D.}~\bibnamefont {Blas}}, \bibinfo
  {author} {\bibfnamefont {E.}~\bibnamefont {Barausse}},  and \bibinfo {author}
  {\bibfnamefont {N.}~\bibnamefont {Yunes}},\ }\href {\doibase
  10.1103/PhysRevD.89.084067} {\bibfield  {journal} {\bibinfo  {journal} {\emph
  {Phys. Rev. D}}\ }\textbf {\bibinfo {volume} {89}},\ \bibinfo {pages}
  {084067} (\bibinfo {year} {2014}{\natexlab{b}})},\ \bibinfo {note} {[Erratum:
  Phys.Rev.D 90, 069902 (2014), Erratum: Phys.Rev.D 90, 069901 (2014)]},\
  \Eprint {http://arxiv.org/abs/1311.7144} {arXiv:1311.7144 [gr-qc]}
  \BibitemShut {NoStop}%
\bibitem [{\citenamefont {Gupta}\ \emph {et~al.}(2021)\citenamefont {Gupta},
  \citenamefont {Herrero-Valea}, \citenamefont {Blas}, \citenamefont
  {Barausse}, \citenamefont {Cornish}, \citenamefont {Yagi},\ and\
  \citenamefont {Yunes}}]{Gupta:2021vdj}%
  \BibitemOpen
  \bibfield  {author} {\bibinfo {author} {\bibfnamefont {T.}~\bibnamefont
  {Gupta}}, \bibinfo {author} {\bibfnamefont {M.}~\bibnamefont
  {Herrero-Valea}}, \bibinfo {author} {\bibfnamefont {D.}~\bibnamefont {Blas}},
  \bibinfo {author} {\bibfnamefont {E.}~\bibnamefont {Barausse}}, \bibinfo
  {author} {\bibfnamefont {N.}~\bibnamefont {Cornish}}, \bibinfo {author}
  {\bibfnamefont {K.}~\bibnamefont {Yagi}},  and \bibinfo {author}
  {\bibfnamefont {N.}~\bibnamefont {Yunes}},\ }\href {\doibase
  10.1088/1361-6382/ac1a69} {\bibfield  {journal} {\bibinfo  {journal} {\emph
  {Class. Quant. Grav.}}\ }\textbf {\bibinfo {volume} {38}},\ \bibinfo {pages}
  {195003} (\bibinfo {year} {2021})},\ \Eprint
  {http://arxiv.org/abs/2104.04596} {arXiv:2104.04596 [gr-qc]} \BibitemShut
  {NoStop}%
\bibitem [{\citenamefont {Hansen}\ \emph {et~al.}(2015)\citenamefont {Hansen},
  \citenamefont {Yunes},\ and\ \citenamefont {Yagi}}]{Hansen:2014ewa}%
  \BibitemOpen
  \bibfield  {author} {\bibinfo {author} {\bibfnamefont {D.}~\bibnamefont
  {Hansen}}, \bibinfo {author} {\bibfnamefont {N.}~\bibnamefont {Yunes}},  and
  \bibinfo {author} {\bibfnamefont {K.}~\bibnamefont {Yagi}},\ }\href {\doibase
  10.1103/PhysRevD.91.082003} {\bibfield  {journal} {\bibinfo  {journal} {\emph
  {Phys. Rev. D}}\ }\textbf {\bibinfo {volume} {91}},\ \bibinfo {pages}
  {082003} (\bibinfo {year} {2015})},\ \Eprint {http://arxiv.org/abs/1412.4132}
  {arXiv:1412.4132 [gr-qc]} \BibitemShut {NoStop}%
\bibitem [{\citenamefont {Zhang}\ \emph
  {et~al.}(2020{\natexlab{a}})\citenamefont {Zhang}, \citenamefont {Zhao},
  \citenamefont {Wang}, \citenamefont {Wang}, \citenamefont {Yagi},
  \citenamefont {Yunes}, \citenamefont {Zhao},\ and\ \citenamefont
  {Zhu}}]{Zhang:2019iim}%
  \BibitemOpen
  \bibfield  {author} {\bibinfo {author} {\bibfnamefont {C.}~\bibnamefont
  {Zhang}}, \bibinfo {author} {\bibfnamefont {X.}~\bibnamefont {Zhao}},
  \bibinfo {author} {\bibfnamefont {A.}~\bibnamefont {Wang}}, \bibinfo {author}
  {\bibfnamefont {B.}~\bibnamefont {Wang}}, \bibinfo {author} {\bibfnamefont
  {K.}~\bibnamefont {Yagi}}, \bibinfo {author} {\bibfnamefont {N.}~\bibnamefont
  {Yunes}}, \bibinfo {author} {\bibfnamefont {W.}~\bibnamefont {Zhao}},  and
  \bibinfo {author} {\bibfnamefont {T.}~\bibnamefont {Zhu}},\ }\href {\doibase
  10.1103/PhysRevD.104.069905} {\bibfield  {journal} {\bibinfo  {journal}
  {\emph {Phys. Rev. D}}\ }\textbf {\bibinfo {volume} {101}},\ \bibinfo {pages}
  {044002} (\bibinfo {year} {2020}{\natexlab{a}})},\ \bibinfo {note} {[Erratum:
  Phys.Rev.D 104, 069905 (2021)]},\ \Eprint {http://arxiv.org/abs/1911.10278}
  {arXiv:1911.10278 [gr-qc]} \BibitemShut {NoStop}%
\bibitem [{\citenamefont {Schumacher}\ \emph {et~al.}(2023)\citenamefont
  {Schumacher}, \citenamefont {Perkins}, \citenamefont {Shaw}, \citenamefont
  {Yagi},\ and\ \citenamefont {Yunes}}]{Schumacher:2023cxh}%
  \BibitemOpen
  \bibfield  {author} {\bibinfo {author} {\bibfnamefont {K.}~\bibnamefont
  {Schumacher}}, \bibinfo {author} {\bibfnamefont {S.~E.}\ \bibnamefont
  {Perkins}}, \bibinfo {author} {\bibfnamefont {A.}~\bibnamefont {Shaw}},
  \bibinfo {author} {\bibfnamefont {K.}~\bibnamefont {Yagi}},  and \bibinfo
  {author} {\bibfnamefont {N.}~\bibnamefont {Yunes}},\ }\href {\doibase
  10.1103/PhysRevD.108.104053} {\bibfield  {journal} {\bibinfo  {journal}
  {\emph {Phys. Rev. D}}\ }\textbf {\bibinfo {volume} {108}},\ \bibinfo {pages}
  {104053} (\bibinfo {year} {2023})},\ \Eprint
  {http://arxiv.org/abs/2304.06801} {arXiv:2304.06801 [gr-qc]} \BibitemShut
  {NoStop}%
\bibitem [{\citenamefont {Eling}\ and\ \citenamefont
  {Jacobson}(2006)}]{Eling:2006ec}%
  \BibitemOpen
  \bibfield  {author} {\bibinfo {author} {\bibfnamefont {C.}~\bibnamefont
  {Eling}} and \bibinfo {author} {\bibfnamefont {T.}~\bibnamefont {Jacobson}},\
  }\href {\doibase 10.1088/0264-9381/23/18/009} {\bibfield  {journal} {\bibinfo
   {journal} {\emph {Class. Quant. Grav.}}\ }\textbf {\bibinfo {volume} {23}},\
  \bibinfo {pages} {5643} (\bibinfo {year} {2006})},\ \bibinfo {note}
  {[Erratum: Class.Quant.Grav. 27, 049802 (2010)]},\ \Eprint
  {http://arxiv.org/abs/gr-qc/0604088} {arXiv:gr-qc/0604088} \BibitemShut
  {NoStop}%
\bibitem [{\citenamefont {Foster}(2006)}]{Foster:2005fr}%
  \BibitemOpen
  \bibfield  {author} {\bibinfo {author} {\bibfnamefont {B.~Z.}\ \bibnamefont
  {Foster}},\ }\href {\doibase 10.1103/PhysRevD.73.024005} {\bibfield
  {journal} {\bibinfo  {journal} {\emph {Phys. Rev. D}}\ }\textbf {\bibinfo
  {volume} {73}},\ \bibinfo {pages} {024005} (\bibinfo {year} {2006})},\
  \Eprint {http://arxiv.org/abs/gr-qc/0509121} {arXiv:gr-qc/0509121}
  \BibitemShut {NoStop}%
\bibitem [{\citenamefont {Konoplya}\ and\ \citenamefont
  {Zhidenko}(2007)}]{Konoplya:2006rv}%
  \BibitemOpen
  \bibfield  {author} {\bibinfo {author} {\bibfnamefont {R.~A.}\ \bibnamefont
  {Konoplya}} and \bibinfo {author} {\bibfnamefont {A.}~\bibnamefont
  {Zhidenko}},\ }\href {\doibase 10.1016/j.physletb.2006.11.036} {\bibfield
  {journal} {\bibinfo  {journal} {\emph {Phys. Lett. B}}\ }\textbf {\bibinfo
  {volume} {644}},\ \bibinfo {pages} {186} (\bibinfo {year} {2007})},\ \Eprint
  {http://arxiv.org/abs/gr-qc/0605082} {arXiv:gr-qc/0605082} \BibitemShut
  {NoStop}%
\bibitem [{\citenamefont {Garfinkle}\ \emph {et~al.}(2007)\citenamefont
  {Garfinkle}, \citenamefont {Eling},\ and\ \citenamefont
  {Jacobson}}]{Garfinkle:2007bk}%
  \BibitemOpen
  \bibfield  {author} {\bibinfo {author} {\bibfnamefont {D.}~\bibnamefont
  {Garfinkle}}, \bibinfo {author} {\bibfnamefont {C.}~\bibnamefont {Eling}},
  and \bibinfo {author} {\bibfnamefont {T.}~\bibnamefont {Jacobson}},\ }\href
  {\doibase 10.1103/PhysRevD.76.024003} {\bibfield  {journal} {\bibinfo
  {journal} {\emph {Phys. Rev. D}}\ }\textbf {\bibinfo {volume} {76}},\
  \bibinfo {pages} {024003} (\bibinfo {year} {2007})},\ \Eprint
  {http://arxiv.org/abs/gr-qc/0703093} {arXiv:gr-qc/0703093} \BibitemShut
  {NoStop}%
\bibitem [{\citenamefont {Berglund}\ \emph {et~al.}(2012)\citenamefont
  {Berglund}, \citenamefont {Bhattacharyya},\ and\ \citenamefont
  {Mattingly}}]{Berglund:2012bu}%
  \BibitemOpen
  \bibfield  {author} {\bibinfo {author} {\bibfnamefont {P.}~\bibnamefont
  {Berglund}}, \bibinfo {author} {\bibfnamefont {J.}~\bibnamefont
  {Bhattacharyya}},  and \bibinfo {author} {\bibfnamefont {D.}~\bibnamefont
  {Mattingly}},\ }\href {\doibase 10.1103/PhysRevD.85.124019} {\bibfield
  {journal} {\bibinfo  {journal} {\emph {Phys. Rev. D}}\ }\textbf {\bibinfo
  {volume} {85}},\ \bibinfo {pages} {124019} (\bibinfo {year} {2012})},\
  \Eprint {http://arxiv.org/abs/1202.4497} {arXiv:1202.4497 [hep-th]}
  \BibitemShut {NoStop}%
\bibitem [{\citenamefont {Gao}\ and\ \citenamefont {Shen}(2013)}]{Gao:2013im}%
  \BibitemOpen
  \bibfield  {author} {\bibinfo {author} {\bibfnamefont {C.}~\bibnamefont
  {Gao}} and \bibinfo {author} {\bibfnamefont {Y.-G.}\ \bibnamefont {Shen}},\
  }\href {\doibase 10.1103/PhysRevD.88.103508} {\bibfield  {journal} {\bibinfo
  {journal} {\emph {Phys. Rev. D}}\ }\textbf {\bibinfo {volume} {88}},\
  \bibinfo {pages} {103508} (\bibinfo {year} {2013})},\ \Eprint
  {http://arxiv.org/abs/1301.7122} {arXiv:1301.7122 [gr-qc]} \BibitemShut
  {NoStop}%
\bibitem [{\citenamefont {Lin}\ \emph {et~al.}(2015)\citenamefont {Lin},
  \citenamefont {Goldoni}, \citenamefont {da~Silva},\ and\ \citenamefont
  {Wang}}]{Lin:2014eaa}%
  \BibitemOpen
  \bibfield  {author} {\bibinfo {author} {\bibfnamefont {K.}~\bibnamefont
  {Lin}}, \bibinfo {author} {\bibfnamefont {O.}~\bibnamefont {Goldoni}},
  \bibinfo {author} {\bibfnamefont {M.~F.}\ \bibnamefont {da~Silva}},  and
  \bibinfo {author} {\bibfnamefont {A.}~\bibnamefont {Wang}},\ }\href {\doibase
  10.1103/PhysRevD.91.024047} {\bibfield  {journal} {\bibinfo  {journal} {\emph
  {Phys. Rev. D}}\ }\textbf {\bibinfo {volume} {91}},\ \bibinfo {pages}
  {024047} (\bibinfo {year} {2015})},\ \Eprint {http://arxiv.org/abs/1410.6678}
  {arXiv:1410.6678 [gr-qc]} \BibitemShut {NoStop}%
\bibitem [{\citenamefont {Ding}\ \emph {et~al.}(2015)\citenamefont {Ding},
  \citenamefont {Wang},\ and\ \citenamefont {Wang}}]{Ding:2015kba}%
  \BibitemOpen
  \bibfield  {author} {\bibinfo {author} {\bibfnamefont {C.}~\bibnamefont
  {Ding}}, \bibinfo {author} {\bibfnamefont {A.}~\bibnamefont {Wang}},  and
  \bibinfo {author} {\bibfnamefont {X.}~\bibnamefont {Wang}},\ }\href {\doibase
  10.1103/PhysRevD.92.084055} {\bibfield  {journal} {\bibinfo  {journal} {\emph
  {Phys. Rev. D}}\ }\textbf {\bibinfo {volume} {92}},\ \bibinfo {pages}
  {084055} (\bibinfo {year} {2015})},\ \Eprint
  {http://arxiv.org/abs/1507.06618} {arXiv:1507.06618 [gr-qc]} \BibitemShut
  {NoStop}%
\bibitem [{\citenamefont {Ding}\ \emph {et~al.}(2016)\citenamefont {Ding},
  \citenamefont {Liu}, \citenamefont {Wang},\ and\ \citenamefont
  {Jing}}]{Ding:2016wcf}%
  \BibitemOpen
  \bibfield  {author} {\bibinfo {author} {\bibfnamefont {C.}~\bibnamefont
  {Ding}}, \bibinfo {author} {\bibfnamefont {C.}~\bibnamefont {Liu}}, \bibinfo
  {author} {\bibfnamefont {A.}~\bibnamefont {Wang}},  and \bibinfo {author}
  {\bibfnamefont {J.}~\bibnamefont {Jing}},\ }\href {\doibase
  10.1103/PhysRevD.94.124034} {\bibfield  {journal} {\bibinfo  {journal} {\emph
  {Phys. Rev. D}}\ }\textbf {\bibinfo {volume} {94}},\ \bibinfo {pages}
  {124034} (\bibinfo {year} {2016})},\ \Eprint
  {http://arxiv.org/abs/1608.00290} {arXiv:1608.00290 [gr-qc]} \BibitemShut
  {NoStop}%
\bibitem [{\citenamefont {Ding}(2017)}]{Ding:2017gfw}%
  \BibitemOpen
  \bibfield  {author} {\bibinfo {author} {\bibfnamefont {C.}~\bibnamefont
  {Ding}},\ }\href {\doibase 10.1103/PhysRevD.96.104021} {\bibfield  {journal}
  {\bibinfo  {journal} {\emph {Phys. Rev. D}}\ }\textbf {\bibinfo {volume}
  {96}},\ \bibinfo {pages} {104021} (\bibinfo {year} {2017})},\ \Eprint
  {http://arxiv.org/abs/1707.06747} {arXiv:1707.06747 [gr-qc]} \BibitemShut
  {NoStop}%
\bibitem [{\citenamefont {Lin}\ \emph {et~al.}(2018)\citenamefont {Lin},
  \citenamefont {Ho},\ and\ \citenamefont {Qian}}]{Lin:2017cmn}%
  \BibitemOpen
  \bibfield  {author} {\bibinfo {author} {\bibfnamefont {K.}~\bibnamefont
  {Lin}}, \bibinfo {author} {\bibfnamefont {F.-H.}\ \bibnamefont {Ho}},  and
  \bibinfo {author} {\bibfnamefont {W.-L.}\ \bibnamefont {Qian}},\ }\href
  {\doibase 10.1142/S0218271819500494} {\bibfield  {journal} {\bibinfo
  {journal} {\emph {Int. J. Mod. Phys. D}}\ }\textbf {\bibinfo {volume} {28}},\
  \bibinfo {pages} {1950049} (\bibinfo {year} {2018})},\ \Eprint
  {http://arxiv.org/abs/1704.06728} {arXiv:1704.06728 [gr-qc]} \BibitemShut
  {NoStop}%
\bibitem [{\citenamefont {Ding}\ and\ \citenamefont
  {Wang}(2019)}]{Ding:2018whp}%
  \BibitemOpen
  \bibfield  {author} {\bibinfo {author} {\bibfnamefont {C.}~\bibnamefont
  {Ding}} and \bibinfo {author} {\bibfnamefont {A.}~\bibnamefont {Wang}},\
  }\href {\doibase 10.1103/PhysRevD.99.124011} {\bibfield  {journal} {\bibinfo
  {journal} {\emph {Phys. Rev. D}}\ }\textbf {\bibinfo {volume} {99}},\
  \bibinfo {pages} {124011} (\bibinfo {year} {2019})},\ \Eprint
  {http://arxiv.org/abs/1811.05779} {arXiv:1811.05779 [gr-qc]} \BibitemShut
  {NoStop}%
\bibitem [{\citenamefont {Chan}\ \emph {et~al.}(2020)\citenamefont {Chan},
  \citenamefont {da~Silva},\ and\ \citenamefont
  {Satheeshkumar}}]{Chan:2019mdn}%
  \BibitemOpen
  \bibfield  {author} {\bibinfo {author} {\bibfnamefont {R.}~\bibnamefont
  {Chan}}, \bibinfo {author} {\bibfnamefont {M.~F.~A.}\ \bibnamefont
  {da~Silva}},  and \bibinfo {author} {\bibfnamefont {V.~H.}\ \bibnamefont
  {Satheeshkumar}},\ }\href {\doibase 10.1088/1475-7516/2020/05/025} {\bibfield
   {journal} {\bibinfo  {journal} {\emph {JCAP}}\ }\textbf {\bibinfo {volume}
  {05}},\ \bibinfo {pages} {025} (\bibinfo {year} {2020})},\ \Eprint
  {http://arxiv.org/abs/1912.12845} {arXiv:1912.12845 [gr-qc]} \BibitemShut
  {NoStop}%
\bibitem [{\citenamefont {Zhu}\ \emph {et~al.}(2019)\citenamefont {Zhu},
  \citenamefont {Wu}, \citenamefont {Jamil},\ and\ \citenamefont
  {Jusufi}}]{Zhu:2019ura}%
  \BibitemOpen
  \bibfield  {author} {\bibinfo {author} {\bibfnamefont {T.}~\bibnamefont
  {Zhu}}, \bibinfo {author} {\bibfnamefont {Q.}~\bibnamefont {Wu}}, \bibinfo
  {author} {\bibfnamefont {M.}~\bibnamefont {Jamil}},  and \bibinfo {author}
  {\bibfnamefont {K.}~\bibnamefont {Jusufi}},\ }\href {\doibase
  10.1103/PhysRevD.100.044055} {\bibfield  {journal} {\bibinfo  {journal}
  {\emph {Phys. Rev. D}}\ }\textbf {\bibinfo {volume} {100}},\ \bibinfo {pages}
  {044055} (\bibinfo {year} {2019})},\ \Eprint
  {http://arxiv.org/abs/1906.05673} {arXiv:1906.05673 [gr-qc]} \BibitemShut
  {NoStop}%
\bibitem [{\citenamefont {Chan}\ \emph {et~al.}(2021)\citenamefont {Chan},
  \citenamefont {Da~Silva},\ and\ \citenamefont
  {Satheeshkumar}}]{Chan:2020amr}%
  \BibitemOpen
  \bibfield  {author} {\bibinfo {author} {\bibfnamefont {R.}~\bibnamefont
  {Chan}}, \bibinfo {author} {\bibfnamefont {M.~F.~A.}\ \bibnamefont
  {Da~Silva}},  and \bibinfo {author} {\bibfnamefont {V.~H.}\ \bibnamefont
  {Satheeshkumar}},\ }\href {\doibase 10.1140/epjc/s10052-021-09120-w}
  {\bibfield  {journal} {\bibinfo  {journal} {\emph {Eur. Phys. J. C}}\
  }\textbf {\bibinfo {volume} {81}},\ \bibinfo {pages} {317} (\bibinfo {year}
  {2021})},\ \Eprint {http://arxiv.org/abs/2003.00227} {arXiv:2003.00227
  [gr-qc]} \BibitemShut {NoStop}%
\bibitem [{\citenamefont {Khodadi}\ and\ \citenamefont
  {Saridakis}(2021)}]{Khodadi:2020gns}%
  \BibitemOpen
  \bibfield  {author} {\bibinfo {author} {\bibfnamefont {M.}~\bibnamefont
  {Khodadi}} and \bibinfo {author} {\bibfnamefont {E.~N.}\ \bibnamefont
  {Saridakis}},\ }\href {\doibase 10.1016/j.dark.2021.100835} {\bibfield
  {journal} {\bibinfo  {journal} {\emph {Phys. Dark Univ.}}\ }\textbf {\bibinfo
  {volume} {32}},\ \bibinfo {pages} {100835} (\bibinfo {year} {2021})},\
  \Eprint {http://arxiv.org/abs/2012.05186} {arXiv:2012.05186 [gr-qc]}
  \BibitemShut {NoStop}%
\bibitem [{\citenamefont {Zhang}\ \emph
  {et~al.}(2020{\natexlab{b}})\citenamefont {Zhang}, \citenamefont {Zhao},
  \citenamefont {Lin}, \citenamefont {Zhang}, \citenamefont {Zhao},\ and\
  \citenamefont {Wang}}]{Zhang:2020too}%
  \BibitemOpen
  \bibfield  {author} {\bibinfo {author} {\bibfnamefont {C.}~\bibnamefont
  {Zhang}}, \bibinfo {author} {\bibfnamefont {X.}~\bibnamefont {Zhao}},
  \bibinfo {author} {\bibfnamefont {K.}~\bibnamefont {Lin}}, \bibinfo {author}
  {\bibfnamefont {S.}~\bibnamefont {Zhang}}, \bibinfo {author} {\bibfnamefont
  {W.}~\bibnamefont {Zhao}},  and \bibinfo {author} {\bibfnamefont
  {A.}~\bibnamefont {Wang}},\ }\href {\doibase 10.1103/PhysRevD.102.064043}
  {\bibfield  {journal} {\bibinfo  {journal} {\emph {Phys. Rev. D}}\ }\textbf
  {\bibinfo {volume} {102}},\ \bibinfo {pages} {064043} (\bibinfo {year}
  {2020}{\natexlab{b}})},\ \Eprint {http://arxiv.org/abs/2004.06155}
  {arXiv:2004.06155 [gr-qc]} \BibitemShut {NoStop}%
\bibitem [{\citenamefont {Oost}\ \emph {et~al.}(2021)\citenamefont {Oost},
  \citenamefont {Mukohyama},\ and\ \citenamefont {Wang}}]{Oost:2021tqi}%
  \BibitemOpen
  \bibfield  {author} {\bibinfo {author} {\bibfnamefont {J.}~\bibnamefont
  {Oost}}, \bibinfo {author} {\bibfnamefont {S.}~\bibnamefont {Mukohyama}},
  and \bibinfo {author} {\bibfnamefont {A.}~\bibnamefont {Wang}},\ }\href
  {\doibase 10.3390/universe7080272} {\bibfield  {journal} {\bibinfo  {journal}
  {\emph {Universe}}\ }\textbf {\bibinfo {volume} {7}},\ \bibinfo {pages} {272}
  (\bibinfo {year} {2021})},\ \Eprint {http://arxiv.org/abs/2106.09044}
  {arXiv:2106.09044 [gr-qc]} \BibitemShut {NoStop}%
\bibitem [{\citenamefont {Mazza}\ and\ \citenamefont
  {Liberati}(2023)}]{Mazza:2023iwv}%
  \BibitemOpen
  \bibfield  {author} {\bibinfo {author} {\bibfnamefont {J.}~\bibnamefont
  {Mazza}} and \bibinfo {author} {\bibfnamefont {S.}~\bibnamefont {Liberati}},\
  }\href {\doibase 10.1007/JHEP03(2023)199} {\bibfield  {journal} {\bibinfo
  {journal} {\emph {JHEP}}\ }\textbf {\bibinfo {volume} {03}},\ \bibinfo
  {pages} {199} (\bibinfo {year} {2023})},\ \Eprint
  {http://arxiv.org/abs/2301.04697} {arXiv:2301.04697 [gr-qc]} \BibitemShut
  {NoStop}%
\bibitem [{\citenamefont {Blas}\ and\ \citenamefont
  {Sibiryakov}(2011)}]{Blas:2011ni}%
  \BibitemOpen
  \bibfield  {author} {\bibinfo {author} {\bibfnamefont {D.}~\bibnamefont
  {Blas}} and \bibinfo {author} {\bibfnamefont {S.}~\bibnamefont
  {Sibiryakov}},\ }\href {\doibase 10.1103/PhysRevD.84.124043} {\bibfield
  {journal} {\bibinfo  {journal} {\emph {Phys. Rev. D}}\ }\textbf {\bibinfo
  {volume} {84}},\ \bibinfo {pages} {124043} (\bibinfo {year} {2011})},\
  \Eprint {http://arxiv.org/abs/1110.2195} {arXiv:1110.2195 [hep-th]}
  \BibitemShut {NoStop}%
\bibitem [{\citenamefont {Mukohyama}\ \emph {et~al.}(2024)\citenamefont
  {Mukohyama}, \citenamefont {Tsujikawa},\ and\ \citenamefont
  {Wang}}]{Mukohyama:2024vsn}%
  \BibitemOpen
  \bibfield  {author} {\bibinfo {author} {\bibfnamefont {S.}~\bibnamefont
  {Mukohyama}}, \bibinfo {author} {\bibfnamefont {S.}~\bibnamefont
  {Tsujikawa}},  and \bibinfo {author} {\bibfnamefont {A.}~\bibnamefont
  {Wang}},\ }\Eprint {http://arxiv.org/abs/2405.14071} {arXiv:2405.14071
  [gr-qc]} \BibitemShut {NoStop}%
\bibitem [{\citenamefont {Tsujikawa}\ \emph {et~al.}(2021)\citenamefont
  {Tsujikawa}, \citenamefont {Zhang}, \citenamefont {Zhao},\ and\ \citenamefont
  {Wang}}]{Tsujikawa:2021typ}%
  \BibitemOpen
  \bibfield  {author} {\bibinfo {author} {\bibfnamefont {S.}~\bibnamefont
  {Tsujikawa}}, \bibinfo {author} {\bibfnamefont {C.}~\bibnamefont {Zhang}},
  \bibinfo {author} {\bibfnamefont {X.}~\bibnamefont {Zhao}},  and \bibinfo
  {author} {\bibfnamefont {A.}~\bibnamefont {Wang}},\ }\href {\doibase
  10.1103/PhysRevD.104.064024} {\bibfield  {journal} {\bibinfo  {journal}
  {\emph {Phys. Rev. D}}\ }\textbf {\bibinfo {volume} {104}},\ \bibinfo {pages}
  {064024} (\bibinfo {year} {2021})},\ \Eprint
  {http://arxiv.org/abs/2107.08061} {arXiv:2107.08061 [gr-qc]} \BibitemShut
  {NoStop}%
\bibitem [{\citenamefont {Regge}\ and\ \citenamefont
  {Wheeler}(1957)}]{Regge:1957td}%
  \BibitemOpen
  \bibfield  {author} {\bibinfo {author} {\bibfnamefont {T.}~\bibnamefont
  {Regge}} and \bibinfo {author} {\bibfnamefont {J.~A.}\ \bibnamefont
  {Wheeler}},\ }\href {\doibase 10.1103/PhysRev.108.1063} {\bibfield  {journal}
  {\bibinfo  {journal} {\emph {Phys. Rev.}}\ }\textbf {\bibinfo {volume}
  {108}},\ \bibinfo {pages} {1063} (\bibinfo {year} {1957})}\BibitemShut
  {NoStop}%
\bibitem [{\citenamefont {Zerilli}(1970)}]{Zerilli:1970se}%
  \BibitemOpen
  \bibfield  {author} {\bibinfo {author} {\bibfnamefont {F.~J.}\ \bibnamefont
  {Zerilli}},\ }\href {\doibase 10.1103/PhysRevLett.24.737} {\bibfield
  {journal} {\bibinfo  {journal} {\emph {Phys. Rev. Lett.}}\ }\textbf {\bibinfo
  {volume} {24}},\ \bibinfo {pages} {737} (\bibinfo {year} {1970})}\BibitemShut
  {NoStop}%
\bibitem [{\citenamefont {De~Felice}\ \emph {et~al.}(2011)\citenamefont
  {De~Felice}, \citenamefont {Suyama},\ and\ \citenamefont
  {Tanaka}}]{DeFelice:2011ka}%
  \BibitemOpen
  \bibfield  {author} {\bibinfo {author} {\bibfnamefont {A.}~\bibnamefont
  {De~Felice}}, \bibinfo {author} {\bibfnamefont {T.}~\bibnamefont {Suyama}},
  and \bibinfo {author} {\bibfnamefont {T.}~\bibnamefont {Tanaka}},\ }\href
  {\doibase 10.1103/PhysRevD.83.104035} {\bibfield  {journal} {\bibinfo
  {journal} {\emph {Phys. Rev. D}}\ }\textbf {\bibinfo {volume} {83}},\
  \bibinfo {pages} {104035} (\bibinfo {year} {2011})},\ \Eprint
  {http://arxiv.org/abs/1102.1521} {arXiv:1102.1521 [gr-qc]} \BibitemShut
  {NoStop}%
\bibitem [{\citenamefont {Kobayashi}\ \emph {et~al.}(2012)\citenamefont
  {Kobayashi}, \citenamefont {Motohashi},\ and\ \citenamefont
  {Suyama}}]{Kobayashi:2012kh}%
  \BibitemOpen
  \bibfield  {author} {\bibinfo {author} {\bibfnamefont {T.}~\bibnamefont
  {Kobayashi}}, \bibinfo {author} {\bibfnamefont {H.}~\bibnamefont
  {Motohashi}},  and \bibinfo {author} {\bibfnamefont {T.}~\bibnamefont
  {Suyama}},\ }\href {\doibase 10.1103/PhysRevD.85.084025} {\bibfield
  {journal} {\bibinfo  {journal} {\emph {Phys. Rev. D}}\ }\textbf {\bibinfo
  {volume} {85}},\ \bibinfo {pages} {084025} (\bibinfo {year} {2012})},\
  \bibinfo {note} {[Erratum: Phys.Rev.D 96, 109903 (2017)]},\ \Eprint
  {http://arxiv.org/abs/1202.4893} {arXiv:1202.4893 [gr-qc]} \BibitemShut
  {NoStop}%
\bibitem [{\citenamefont {Kobayashi}\ \emph {et~al.}(2014)\citenamefont
  {Kobayashi}, \citenamefont {Motohashi},\ and\ \citenamefont
  {Suyama}}]{Kobayashi:2014wsa}%
  \BibitemOpen
  \bibfield  {author} {\bibinfo {author} {\bibfnamefont {T.}~\bibnamefont
  {Kobayashi}}, \bibinfo {author} {\bibfnamefont {H.}~\bibnamefont
  {Motohashi}},  and \bibinfo {author} {\bibfnamefont {T.}~\bibnamefont
  {Suyama}},\ }\href {\doibase 10.1103/PhysRevD.89.084042} {\bibfield
  {journal} {\bibinfo  {journal} {\emph {Phys. Rev. D}}\ }\textbf {\bibinfo
  {volume} {89}},\ \bibinfo {pages} {084042} (\bibinfo {year} {2014})},\
  \Eprint {http://arxiv.org/abs/1402.6740} {arXiv:1402.6740 [gr-qc]}
  \BibitemShut {NoStop}%
\bibitem [{\citenamefont {Kase}\ and\ \citenamefont
  {Tsujikawa}(2022)}]{Kase:2021mix}%
  \BibitemOpen
  \bibfield  {author} {\bibinfo {author} {\bibfnamefont {R.}~\bibnamefont
  {Kase}} and \bibinfo {author} {\bibfnamefont {S.}~\bibnamefont {Tsujikawa}},\
  }\href {\doibase 10.1103/PhysRevD.105.024059} {\bibfield  {journal} {\bibinfo
   {journal} {\emph {Phys. Rev. D}}\ }\textbf {\bibinfo {volume} {105}},\
  \bibinfo {pages} {024059} (\bibinfo {year} {2022})},\ \Eprint
  {http://arxiv.org/abs/2110.12728} {arXiv:2110.12728 [gr-qc]} \BibitemShut
  {NoStop}%
\bibitem [{\citenamefont {De~Felice}\ and\ \citenamefont
  {Tsujikawa}(2024)}]{DeFelice:2023rra}%
  \BibitemOpen
  \bibfield  {author} {\bibinfo {author} {\bibfnamefont {A.}~\bibnamefont
  {De~Felice}} and \bibinfo {author} {\bibfnamefont {S.}~\bibnamefont
  {Tsujikawa}},\ }\href {\doibase 10.1103/PhysRevD.109.084022} {\bibfield
  {journal} {\bibinfo  {journal} {\emph {Phys. Rev. D}}\ }\textbf {\bibinfo
  {volume} {109}},\ \bibinfo {pages} {084022} (\bibinfo {year} {2024})},\
  \Eprint {http://arxiv.org/abs/2312.03191} {arXiv:2312.03191 [gr-qc]}
  \BibitemShut {NoStop}%
\bibitem [{\citenamefont {Kase}\ and\ \citenamefont
  {Tsujikawa}(2023)}]{Kase:2023kvq}%
  \BibitemOpen
  \bibfield  {author} {\bibinfo {author} {\bibfnamefont {R.}~\bibnamefont
  {Kase}} and \bibinfo {author} {\bibfnamefont {S.}~\bibnamefont {Tsujikawa}},\
  }\href {\doibase 10.1103/PhysRevD.107.104045} {\bibfield  {journal} {\bibinfo
   {journal} {\emph {Phys. Rev. D}}\ }\textbf {\bibinfo {volume} {107}},\
  \bibinfo {pages} {104045} (\bibinfo {year} {2023})},\ \Eprint
  {http://arxiv.org/abs/2301.10362} {arXiv:2301.10362 [gr-qc]} \BibitemShut
  {NoStop}%
\bibitem [{\citenamefont {Chen}\ \emph {et~al.}(2024)\citenamefont {Chen},
  \citenamefont {De~Felice},\ and\ \citenamefont {Tsujikawa}}]{Chen:2024hkm}%
  \BibitemOpen
  \bibfield  {author} {\bibinfo {author} {\bibfnamefont {C.-Y.}\ \bibnamefont
  {Chen}}, \bibinfo {author} {\bibfnamefont {A.}~\bibnamefont {De~Felice}},
  and \bibinfo {author} {\bibfnamefont {S.}~\bibnamefont {Tsujikawa}},\
  }\Eprint {http://arxiv.org/abs/2404.09377} {arXiv:2404.09377 [gr-qc]}
  \BibitemShut {NoStop}%
\bibitem [{\citenamefont {De~Felice}\ and\ \citenamefont
  {Tsujikawa}(2023)}]{DeFelice:2023kpl}%
  \BibitemOpen
  \bibfield  {author} {\bibinfo {author} {\bibfnamefont {A.}~\bibnamefont
  {De~Felice}} and \bibinfo {author} {\bibfnamefont {S.}~\bibnamefont
  {Tsujikawa}},\ }\href {\doibase 10.1088/1475-7516/2023/10/004} {\bibfield
  {journal} {\bibinfo  {journal} {\emph {JCAP}}\ }\textbf {\bibinfo {volume}
  {10}},\ \bibinfo {pages} {004} (\bibinfo {year} {2023})},\ \Eprint
  {http://arxiv.org/abs/2307.06490} {arXiv:2307.06490 [gr-qc]} \BibitemShut
  {NoStop}%
\bibitem [{\citenamefont {Tasinato}(2014)}]{Tasinato:2014eka}%
  \BibitemOpen
  \bibfield  {author} {\bibinfo {author} {\bibfnamefont {G.}~\bibnamefont
  {Tasinato}},\ }\href {\doibase 10.1007/JHEP04(2014)067} {\bibfield  {journal}
  {\bibinfo  {journal} {\emph {JHEP}}\ }\textbf {\bibinfo {volume} {04}},\
  \bibinfo {pages} {067} (\bibinfo {year} {2014})},\ \Eprint
  {http://arxiv.org/abs/1402.6450} {arXiv:1402.6450 [hep-th]} \BibitemShut
  {NoStop}%
\bibitem [{\citenamefont {Heisenberg}(2014)}]{Heisenberg:2014rta}%
  \BibitemOpen
  \bibfield  {author} {\bibinfo {author} {\bibfnamefont {L.}~\bibnamefont
  {Heisenberg}},\ }\href {\doibase 10.1088/1475-7516/2014/05/015} {\bibfield
  {journal} {\bibinfo  {journal} {\emph {JCAP}}\ }\textbf {\bibinfo {volume}
  {05}},\ \bibinfo {pages} {015} (\bibinfo {year} {2014})},\ \Eprint
  {http://arxiv.org/abs/1402.7026} {arXiv:1402.7026 [hep-th]} \BibitemShut
  {NoStop}%
\bibitem [{\citenamefont {Kubota}\ \emph {et~al.}(2023)\citenamefont {Kubota},
  \citenamefont {Arai},\ and\ \citenamefont {Mukohyama}}]{Kubota:2022lbn}%
  \BibitemOpen
  \bibfield  {author} {\bibinfo {author} {\bibfnamefont {K.-i.}\ \bibnamefont
  {Kubota}}, \bibinfo {author} {\bibfnamefont {S.}~\bibnamefont {Arai}},  and
  \bibinfo {author} {\bibfnamefont {S.}~\bibnamefont {Mukohyama}},\ }\href
  {\doibase 10.1103/PhysRevD.107.064002} {\bibfield  {journal} {\bibinfo
  {journal} {\emph {Phys. Rev. D}}\ }\textbf {\bibinfo {volume} {107}},\
  \bibinfo {pages} {064002} (\bibinfo {year} {2023})},\ \Eprint
  {http://arxiv.org/abs/2209.00795} {arXiv:2209.00795 [gr-qc]} \BibitemShut
  {NoStop}%
\end{thebibliography}%

\end{document}